\documentclass[aps,prb,reprint,nolongbibliography,noeprint,superscriptaddress]{revtex4-2}
\usepackage{amsfonts,amssymb,amsmath,bm,graphicx,color}
\usepackage[colorlinks=true,citecolor=blue,linkcolor=blue,urlcolor=blue]{hyperref}
\allowdisplaybreaks[4]
\pdfmapline{=cmr12 CMR12 <cmr12.pfb <f7b6d320.enc}
\newcommand\as{{\rm \AA}}
\newcommand\ev{{\rm eV}}
\newcommand\mev{{\rm meV}}
\newcommand\hb[1]{\hat {\bm #1}}
\newcommand\hc[1]{\hat {\cal #1}}

\begin{document}
%--- Front Matter
\title{Magnetization energy current in the axial magnetic effect}
\author{Atsuo Shitade}
\affiliation{Institute for Molecular Science, Aichi 444-8585, Japan}
\author{Yasufumi Araki}
\affiliation{Advanced Science Research Center, Japan Atomic Energy Agency (JAEA), Tokai 319-1195, Japan}
\date{\today}
\begin{abstract}
  The axial magnetic effect (AME) is one of the anomalous transport phenomena in which the energy current is induced by an axial magnetic field.
  Here, we numerically study the AME for the relativistic Wilson fermion in the axial magnetic field and a twisted Dirac semimetal.
  The AME current density inside the bulk is nonzero,
  and particularly in the low-energy regime for the former model, it is explained by the field-theoretical results without any fitting parameter.
  However, for both models, the average AME current density vanishes owing to the surface contribution.
  The axial gauge field is regarded as the spatially modulated (effective) Zeeman field and induces the spatially modulated energy magnetization.
  The AME is attributed to the magnetization energy current and hence cannot be observed in transport experiments.
\end{abstract}
\maketitle
%--- Main Matter
\section{Introduction} \label{sec:introduction}
The chiral anomaly has attracted much attention from the high-energy and condensed matter communities.
Chiral fermions have the chiral symmetry in the classical action
but not in the quantum-mechanical partition function when parallel electromagnetic fields are applied~\cite{PhysRev.177.2426,Bell1969}.
Such systems are realized in quark-gluon plasmas in heavy-ion collision experiments~\cite{ARSENE20051,BACK200528,ADAMS2005102,ADCOX2005184}
and effectively in Dirac and Weyl semimetals~\cite{RevModPhys.90.015001}.

The chiral anomaly gives rise to various anomalous transport phenomena.
Among them is the chiral magnetic effect (CME),
in which the charge current is induced by a magnetic field~\cite{PhysRevD.22.3080,KHARZEEV2006260,KHARZEEV200767,KHARZEEV2008227,PhysRevD.78.074033}.
In the relativistic case, it is expressed as
\begin{subequations} \begin{align}
  {\bm j}_{\chi}
  = & q^2 \chi \mu_{\chi} {\bm B}_{\chi}/4 \pi^2 \hbar^2, \label{eq:cme1a} \\
  {\bm j}
  = & {\bm j}_{\rm R} + {\bm j}_{\rm L}
  = q^2 (\mu_5 {\bm B} + \mu {\bm B}_5)/2 \pi^2 \hbar^2, \label{eq:cme1b} \\
  {\bm j}_5
  = & {\bm j}_{\rm R} - {\bm j}_{\rm L}
  = q^2 (\mu {\bm B} + \mu_5 {\bm B}_5)/2 \pi^2 \hbar^2. \label{eq:cme1c}
\end{align} \label{eq:cme1}\end{subequations}
Here, ${\bm j}_{\rm R/L}$, $\mu_{\rm R/L}$, and ${\bm B}_{\rm R/L}$ are the charge current density, chemical potential, and magnetic field for right- and left-handed fermions, respectively.
$q$ is the electric charge, $\mu (\mu_5) = (\mu_{\rm R} \pm \mu_{\rm L})/2$ is the (chiral) chemical potential,
and ${\bm B} ({\bm B}_5) = ({\bm B}_{\rm R} \pm {\bm B}_{\rm L})/2$ is an (axial) magnetic field.
The first term of Eq.~\eqref{eq:cme1b} is the CME in a narrow sense, while the second term is called the chiral pseudomagnetic effect (CPME)~\cite{PhysRevX.6.041021,PhysRevX.6.041046}.
In equilibrium, the CME does not occur because nonzero $\mu_5$ cannot be realized~\cite{Zhou_2013,PhysRevLett.111.027201,Landsteiner2016,PhysRevLett.118.127601},
as forbidden by the Bloch-Bohm theorem~\cite{PhysRev.75.502,JPSJ.65.3254,PhysRevD.92.085011}.
By applying parallel electric and magnetic fields,
the chiral imbalance is generated away from equilibrium and results in the negative magnetoresistance via the CME~\cite{NIELSEN1983389,PhysRevB.88.104412}.
In the condensed matter context, the phenomenon was experimentally observed
in a noncentrosymmetric Weyl semimetal TaAs and its family~\cite{PhysRevx.5.031023,Du2016,Zhang2016,PhysRevB.93.121112}.

Another anomalous transport phenomenon is the chiral vortical effect (CVE),
in which the charge current is induced by the vorticity~\cite{PhysRevD.20.1807,PhysRevD.21.2260,KHARZEEV200767,Erdmenger2009,Banerjee2011,PhysRevLett.103.191601,PhysRevLett.107.021601,Landsteiner2011,PhysRevLett.113.182302}.
In the relativistic case, it is expressed as
\begin{subequations} \begin{align}
  {\bm j}_{\chi}
  = & q \chi (\mu_{\chi}^2 + \pi^2 T^2/3) {\bm \omega}/4 \pi^2 \hbar^2, \label{eq:cme2a} \\
  {\bm j}
  = & q \mu \mu_5 {\bm \omega}/\pi^2 \hbar^2, \label{eq:cme2b} \\
  {\bm j}_5
  = & q (\mu^2 + \mu_5^2 + \pi^2 T^2/3) {\bm \omega}/2 \pi^2 \hbar^2. \label{eq:cme2c}
\end{align} \label{eq:cme2}\end{subequations}
Here, $T$ is the temperature, and ${\bm \omega}$ is the vorticity.
Since the rotating system of chiral fermions is in equilibrium, the CVE current is not expected to flow.
However, the Bloch-Bohm theorem cannot be applied to the CVE, because the theorem is valid only in the thermodynamic limit,
while the system size is limited by the causality~\cite{PhysRevD.21.2260,PhysRevD.92.085011}.
Recently, we found that the transport current of the CVE vanishes regardless of the presence or absence of $\mu_5$~\cite{PhysRevB.102.205201}.
In other words, the local charge current of the CVE is just the magnetization charge current that cannot be observed in transport experiments.
We also demonstrated that the anisotropic CVE can be observed in condensed matter systems that belong to some chiral point groups.

The axial magnetic effect (AME) is also an anomalous transport phenomenon in which the energy current is induced by an axial magnetic field~\cite{PhysRevD.88.071501,Buividovich_2015,PhysRevB.89.081407}.
In the relativistic case, the energy-momentum tensor is symmetric, and the AME is reciprocal to the CVE in Eq.~\eqref{eq:cme2c}.
Hence it is expressed as
\begin{equation}
  {\bm j}_{\varepsilon}
  = q (\mu^2 + \mu_5^2 + \pi^2 T^2/3) {\bm B}_5/4 \pi^2 \hbar^2. \label{eq:cme3}
\end{equation}
The nonzero temperature part of Eq.~\eqref{eq:cme3}, $\pi^2 T^2/3$, was numerically supported by the large-scale SU($2$) lattice gauge theory~\cite{Buividovich_2015}.
However, since the transport current of the CVE vanishes~\cite{PhysRevB.102.205201}, that of the AME should vanish as well.
More generally, it is natural to expect that the energy current should not flow in equilibrium,
although the Bloch-Bohm theorem for the energy current has not been proved yet.

An axial gauge field can be effectively engineered
in Dirac and Weyl semimetals~\cite{PhysRevB.87.235306,*PhysRevB.92.119904,PhysRevLett.115.177202,PhysRevX.6.041021,PhysRevX.6.041046,PhysRevB.94.115312}.
In magnetic Weyl semimetals, the Zeeman field from magnetic moments acts as the axial gauge field,
and a magnetic texture yields an axial magnetic field~\cite{PhysRevB.87.235306,*PhysRevB.92.119904,PhysRevX.6.041046,PhysRevB.94.115312}.
Elastic strain can also generate the axial gauge field in magnetic Weyl semimetals~\cite{PhysRevLett.115.177202}
and nonmagnetic Dirac and Weyl semimetals~\cite{PhysRevX.6.041021,PhysRevX.6.041046}.
Note that the axial gauge field in the latter case is spin dependent, because strain preserves the time-reversal symmetry.
Thus magnetic or strained Dirac and Weyl semimetals are ideal platforms to test the possibility for the above anomalous transport phenomena including the AME~\cite{Ilan2019}.

In this paper, we numerically show that the AME is canceled by the surface contribution.
We consider two different models; one is the relativistic Wilson fermion~\cite{PhysRevD.10.2445} in an axial magnetic field,
and the other is a lattice model of a twisted Dirac semimetal Cd$_3$As$_2$~\cite{PhysRevX.6.041021}.
We calculate the charge and energy current densities imposing the open boundary conditions.
Our purpose for considering the former model is to compare our numerical results with the field-theoretical ones in the high-energy literature.
Indeed, in the low-energy regime, we find that the charge and energy current densities inside the bulk are explained by Eqs.~\eqref{eq:cme1b} and \eqref{eq:cme3} without any fitting parameter.
For both models, the current densities inside the bulk are nonzero, but the average ones vanish owing to the surface contributions.
Such current distributions are explained by the magnetization charge and energy currents,
namely, the spatially modulated orbital and energy magnetizations induced by the spatially modulated (effective) Zeeman field.
In equilibrium, neither the CPME nor the AME can be observed in transport experiments.

Let us mention the previous results on the CPME~\cite{PhysRevX.6.041046} and AME~\cite{PhysRevB.89.081407} in the condensed matter context.
In Ref.~\cite{PhysRevX.6.041046}, it was numerically shown that the average charge current density vanishes owing to the surface contribution.
Thus it is natural to expect that the average energy current density, on which we mainly focus, vanishes as well.
In Ref.~\cite{PhysRevB.89.081407}, the authors considered a magnetic Weyl semimetal where the azimuthal axial magnetic field exists only at the surface.
In this setup, by construction, the resulting energy current is the magnetization current circulating at the surface.
However, discussion was lacking on a more generic setup where the axial magnetic field exists not only at the surface but also inside the bulk.
We give a solid explanation for the vanishing of  the average charge and energy current densities by explicitly calculating the orbital and energy magnetizations.
We also discuss the difference between the energy magnetization and the orbital angular momentum in condensed matter systems.

\section{Wilson fermion} \label{sec:wilson}
The relativistic Wilson fermion is expressed as~\cite{PhysRevD.10.2445}
\begin{equation}
  {\hc H}_0({\bm k})
  = {\bm h}({\bm k}) \cdot \rho_z {\bm \sigma} + m({\bm k}) \rho_x, \label{eq:wilson1}
\end{equation}
in which $h^x({\bm k}) = v \sin k_x a$, $h^y({\bm k}) = v \sin k_y a$, $h^z({\bm k}) = v \sin k_z a$,
and $m({\bm k}) = m + r (3 - \cos k_x a - \cos k_y a - \cos k_z a)$ is the Wilson term.
$\rho$ and $\sigma$ are the Pauli matrices for the particle-hole and spin degrees of freedom, respectively,
and form the Dirac representation of the Dirac matrices as ${\bm \alpha} = \rho_z {\bm \sigma}$ and $\beta = \rho_x$.
Below, we choose $v = 1$ as the energy unit and $r = (2/\sqrt{3}) v$.
The lattice constant is $a = 1$.

We introduce an axial gauge field in the Landau gauge, i.e.,
\begin{equation}
  {\hc H}_1(x)
  = - (q v a/\hbar) A_{5y}(x) \sigma_y, \label{eq:wilson2}
\end{equation}
with $A_{5y}(x) = B_5^z x$.
Note that such a way of introducing the axial gauge field is different from the conventional way of introducing the vector gauge field using the Wilson line.
Since the Wilson term breaks the chiral symmetry, it is impossible to introduce the axial gauge field in a gauge-invariant way.
We choose $q B_5^z a^2/\hbar = 1 \times 10^{-3}$.

\begin{figure}
  \centering
  \includegraphics[clip,width=0.48\textwidth]{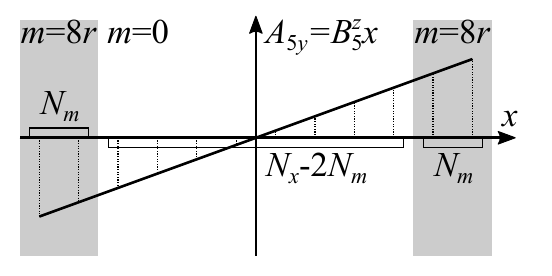}
  \caption{%
  Configuration of the mass $m$ and axial gauge field $A_{5y}$.
  The massless region of $N_x - 2 N_m$ sites is sandwiched by the shaded massive regions of $N_m$ sites.%
  } \label{fig:wilson_setup}
\end{figure}

\begin{figure}
  \centering
  \includegraphics[clip,width=0.48\textwidth]{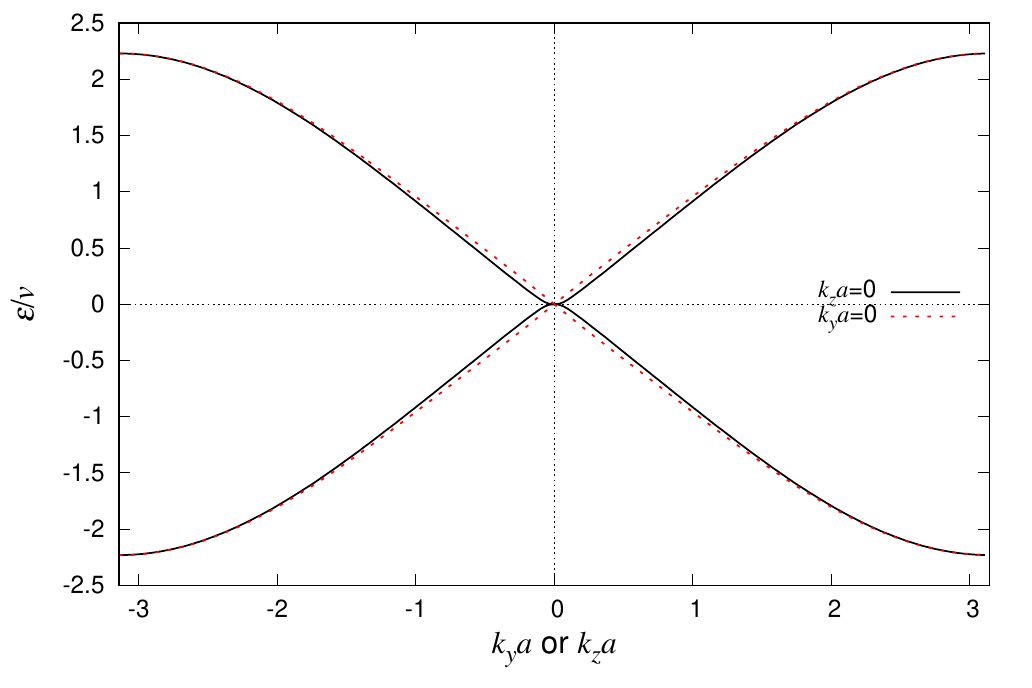}
  \caption{%
  Lowest positive and highest negative eigenvalues of the total Hamiltonian [Eq.~\eqref{eq:wilson1} $+$ Eq.~\eqref{eq:wilson2}] for the Wilson fermion.
  Black solid and red dashed lines represent the eigenvalues as functions of $k_y a$ ($k_z a = 0$) and $k_z a$ ($k_y a = 0$), respectively.%
  } \label{fig:wilson_band}
\end{figure}

We diagonalize the total Hamiltonian [Eq.~\eqref{eq:wilson1} $+$ Eq.~\eqref{eq:wilson2}] for the setup depicted in Fig.~\ref{fig:wilson_setup}.
The mass depends on $x$; $m = 0$ in the massless region of $N_x - 2 N_m$ sites, while $m = 8 r$ in the massive regions of $N_m$ sites.
The axial gauge field is present in the massless and massive regions continuously.
We impose the open boundary condition in the $x$ axis and the periodic ones in the $y$ and $z$ axes.
The numbers of sites are $N_x = 300$ and $N_m = 50$, leading to $N_x - 2 N_m = 200$, and $N_y = N_z = 192$.
Figure~\ref{fig:wilson_band} shows the lowest positive eigenvalue for particles and the highest negative one for antiparticles.
In the low-energy regime, the eigenvalues are quadratic as functions of $k_y a$ ($k_z a = 0$) and linear as functions of $k_z a$ ($k_y a = 0$),
which is consistent with the chiral Landau levels in the presence of the Wilson term.

\begin{figure*}
  \centering
  \includegraphics[clip,width=0.98\textwidth]{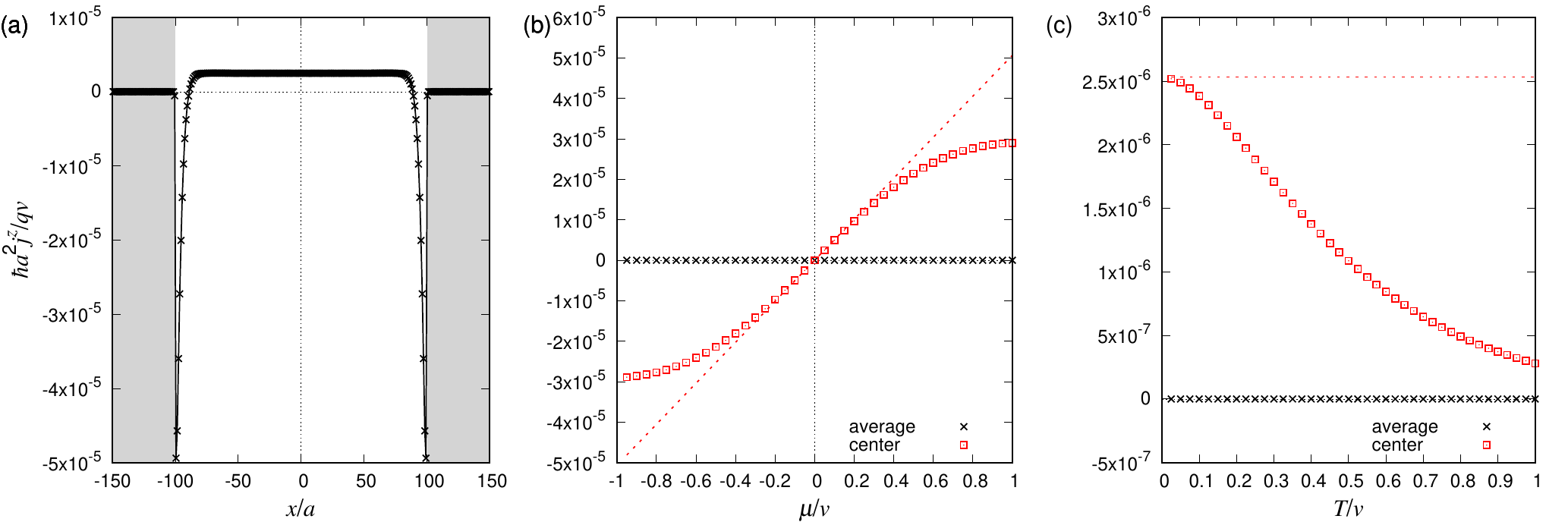}
  \caption{%
  (a) Spatial dependence of the charge current density for the Wilson fermion in the axial magnetic field at $\mu/v = T/v = 0.05$.
  The shaded areas represent the massive regions.
  (b) The chemical potential dependence at $T/v = 0.05$ and (c) the temperature dependence at $\mu/v = 0.05$
  of the average charge current density (black crosses) and that at the center (red squares).
  The red dashed lines represent the second term of Eq.~\eqref{eq:cme1b} with $q B_5^z a^2/\hbar = 1 \times 10^{-3}$.%
  } \label{fig:wilson_cme}
\end{figure*}

\begin{figure*}
  \centering
  \includegraphics[clip,width=0.98\textwidth]{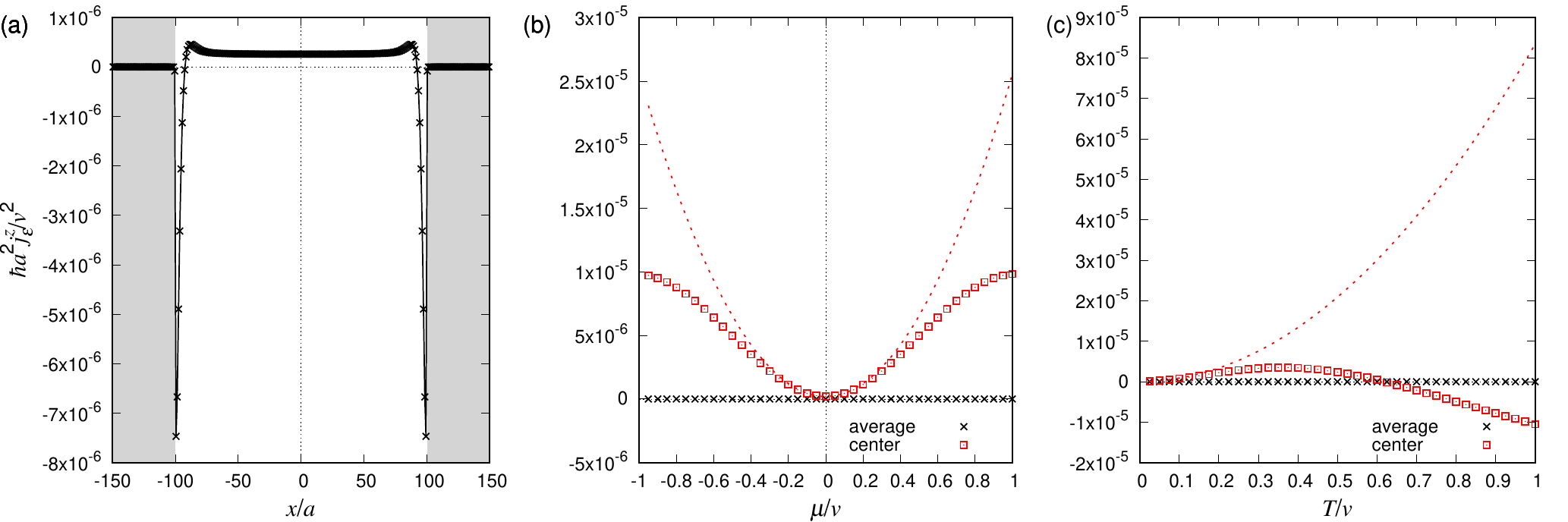}
  \caption{%
  (a) Spatial dependence of the energy current density for the Wilson fermion in the axial magnetic field at $\mu/v = T/v = 0.05$.
  The shaded areas represent the massive regions.
  (b) The chemical potential dependence at $T/v = 0.05$ and (c) the temperature dependence at $\mu/v = 0.05$
  of the average energy current density (black crosses) and that at the center (red squares).
  The red dashed lines represent Eq.~\eqref{eq:cme3} with $q B_5^z a^2/\hbar = 1 \times 10^{-3}$.%
  } \label{fig:wilson_ame}
\end{figure*}

The charge and energy current densities are
\begin{subequations} \begin{align}
  j^z(x)
  = & \frac{q}{a^3 N_y N_z} \sum_{n k_y k_z} v_n^z(x, k_y, k_z) f(\epsilon_n(k_y, k_z)), \label{eq:wilson3a} \\
  j_{\varepsilon}^z(x)
  = & \frac{1}{a^3 N_y N_z} \sum_{n k_y k_z} v_n^z(x, k_y, k_z) \epsilon_n(k_y, k_z) \notag \\
  & \times f(\epsilon_n(k_y, k_z)), \label{eq:wilson3b}
\end{align} \label{eq:wilson3}\end{subequations}
where $v_n^z(x, k_y, k_z) = \hbar^{-1} \langle u_n(k_y, k_z) | \partial_{k_z} {\hc H}(x, k_y, k_z) | u_n(k_y, k_z) \rangle$.
The distribution function is $f(\epsilon) = [e^{(\epsilon - \mu)/T} + 1]^{-1}$ for $\epsilon > 0$ and $f(\epsilon) = -[e^{(|\epsilon| + \mu)/T} + 1]^{-1}$ for $\epsilon < 0$.
The average current densities are obtained by
\begin{equation}
  j_{(\varepsilon) {\rm ave}}^z
  = \frac{1}{N_x - 2 N_m} \sum_x j_{(\varepsilon)}^z(x). \label{eq:wilson4}
\end{equation}
In Fig.~\ref{fig:wilson_cme}(a), the charge current density at the surface is opposite to that inside the bulk.
In Figs.~\ref{fig:wilson_cme}(b) and \ref{fig:wilson_cme}(c), the charge current density inside the bulk is completely explained by the second term of Eq.~\eqref{eq:cme1b} without any fitting parameter,
as far as the low-energy regime ($|\mu/v| < 0.4$ and $T/v < 0.05$) is concerned.
However, the average charge current density vanishes for any chemical potential $\mu$ and temperature $T$~\cite{PhysRevX.6.041046}.
The same is true for the energy current density as shown in Fig.~\ref{fig:wilson_ame}.
We conclude that Eqs.~\eqref{eq:cme1b} and \eqref{eq:cme3} correctly describe the CPME and AME inside the bulk, respectively,
but are completely canceled by the surface contributions.

Let us comment on the previous numerical results on the CPME in Ref.~\cite{PhysRevX.6.041046}.
The authors considered a similar model of a Weyl semimetal using $A_{5y}(x) = A_{5y} + B_5^z x$ in Eq.~\eqref{eq:wilson2}.
Hence the spatial dependence of the charge current density is different from ours, but the average one vanishes as expected.
They also showed the charge current density at the center of the system but did not relate it to Eq.~\eqref{eq:cme1b}.

We also comment on other previous numerical results on the AME using the lattice gauge theory~\cite{PhysRevD.88.071501,Buividovich_2015}.
The authors used the overlap fermion that preserves the modified chiral symmetry~\cite{NEUBERGER1998141}.
They obtained the nonzero temperature part of Eq.~\eqref{eq:cme3} only, because the open boundary condition cannot be imposed in their setup.
In our setup, the chiral symmetry is broken in the high-energy regime,
which causes the deviation of the charge and energy current densities inside the bulk from the expected values.
In the low-energy regime, our results are consistent with the previous ones.

\section{Lattice model of Cd$_3$As$_2$} \label{sec:Cd3As2}
Cd$_3$As$_2$ was predicted to be a Dirac semimetal by first-principles calculation~\cite{PhysRevB.88.125427}
and soon it was experimentally confirmed~\cite{Neupane2014,Liu2014,PhysRevLett.113.027603,Jeon2014}.
The effective model near the $\Gamma$ point is expressed as~\cite{PhysRevX.6.041021}
\begin{equation}
  {\hc H}_0({\bm k})
  = h^0({\bm k}) + {\bm h}({\bm k}) \cdot {\bm \rho} \sigma_z. \label{eq:Cd3As21}
\end{equation}
Here, $h^0({\bm k}) = c_0 + (2 c_1/c^2) (1 - \cos k_z c) + (2 c_2/a^2) (2 - \cos k_x a - \cos k_y a)$,
$h^x({\bm k}) = (v/a) \sin k_x a$, $h^y({\bm k}) = (v/a) \sin k_y a$,
and $h^z({\bm k}) = m_0 - (2 m_1/c^2) (1 - \cos k_z c) - (2 m_2/a^2) (2 - \cos k_x a - \cos k_y a)$.
Two sets of the Pauli matrices, $\rho$ and $\sigma$, represent the spin-orbit coupled states $| 3/2, \pm 3/2 \rangle$ and $| 1/2, \pm 1/2 \rangle$.
Since two sectors of $\sigma_z = \pm 1$ are decoupled in Eq.~\eqref{eq:Cd3As21},
below we focus on the $\sigma_z = +1$ sector only and call the model $1/2$-Cd$_3$As$_2$ following Ref.~\cite{PhysRevX.6.041021}.
The material parameters are $c_0 = -0.0145~\ev$, $c_1 = 10.59~\ev \as^2$, $c_2 = 11.5~\ev \as^2$,
$m_0 = 0.0205~\ev$, $m_1 = 18.77~\ev \as^2$, $m_2 = 13.5~\ev \as^2$, and $v = 0.889~\ev \as$~\cite{PhysRevB.95.161306}.
The lattice constants are $a = 12.67~\as$ and $c = 25.48~\as$.

The Hamiltonian~\eqref{eq:Cd3As21} is gapless at ${\bm k} = \mp {\bm Q} = [0, 0, \mp Q]^{\rm T}$
with $Q$ satisfying $m_0 = (2 m_1/c^2) (1 - \cos Q c)$.
Around these points, Eq.~\eqref{eq:Cd3As21} is approximated as
\begin{equation}
  {\hc H}_{0 \pm}(\delta {\bm k})
  = {\tilde c}_0 \mp r v_0 \delta k_z + v \delta {\bm k}_{\perp} \cdot {\bm \rho}_{\perp} \pm r v \delta k_z \rho_z, \label{eq:Cd3As22}
\end{equation}
where ${\tilde c}_0 = c_0 + m_0 c_1/m_1$, $v_0 = v c_1/m_1$, and $r = (2 m_1/v c) \sin Q c$.
Using the above parameters, we obtain $Q c = 0.869$, ${\tilde c}_0 = -0.00293~\ev$, $v_0 = 0.502~\ev \as$, and $r = 1.266$.
Thus, an anisotropic Weyl semimetal is effectively realized in $1/2$-Cd$_3$As$_2$.

Twisting $1/2$-Cd$_3$As$_2$ effectively generates a uniform axial magnetic field~\cite{PhysRevX.6.041021}.
Displacement is expressed as ${\bm u}({\bm x}) = \theta (z/L_z) {\bm x} \times {\bm e}_z$,
where $\theta$ is the twist angle and $L_z$ is the length in the $z$ axis.
Strain $u_{ij} = (\partial_{x^i} u_j + \partial_{x^j} u_i)/2$ is then expressed as $u_{xz}({\bm x}_{\perp}) = \theta y/2 L_z$ and $u_{yz}({\bm x}_{\perp}) = -\theta x/2 L_z$
and coupled to electrons as
\begin{align}
  {\hc H}_1({\bm x}_{\perp}, k_z)
  = & g (v/a) {\bm u}_{\perp z}({\bm x}_{\perp}) \cdot {\bm \rho}_{\perp} \sin k_z c \notag \\
  = & \frac{q v}{\hbar} \frac{\sin k_z c}{\sin Q c} {\bm A}_{5 \perp}({\bm x}_{\perp}) \cdot {\bm \rho}_{\perp}. \label{eq:Cd3As23}
\end{align}
Here, $g$ is a dimensionless coupling constant, ${\bm A}_{5 \perp}({\bm x}_{\perp}) = B_5^z [-y, x]^{\rm T}/2$, and $B_5^z = -(\hbar/q a L_z) g \theta \sin Q c$.
Around the Weyl points, the total Hamiltonian is approximated as
\begin{align}
  {\hc H}_{\pm}({\bm x}_{\perp}, \delta k_z)
  = & {\tilde c}_0 \mp r v_0 \delta k_z + v [\delta {\hb k}_{\perp} \mp q {\bm A}_{5 \perp}({\bm x}_{\perp})/\hbar] \cdot {\bm \rho}_{\perp} \notag \\
  & \pm r v \delta k_z \rho_z. \label{eq:Cd3As24}
\end{align}
Thus ${\bm A}_5$ and ${\bm B}_5$ effectively act as the axial gauge and magnetic fields, respectively.
Since twist preserves the time-reversal symmetry,
${\bm A}_5$ and ${\bm B}_5$ depend on the spin if both sectors of $\sigma_z = \pm 1$ are considered.
In this case, the charge and energy currents below should read the spin current and spin-dependent energy current, respectively.

\begin{figure}
  \centering
  \includegraphics[clip,width=0.48\textwidth]{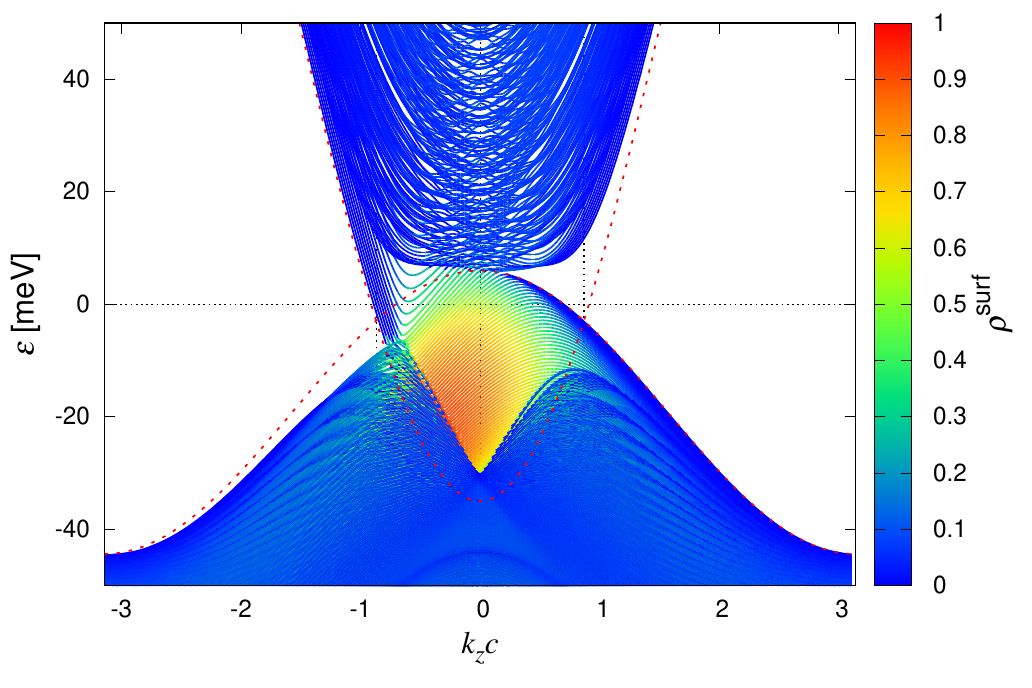}
  \caption{%
  Band structure of the total Hamiltonian [Eq.~\eqref{eq:Cd3As21} $+$ Eq.~\eqref{eq:Cd3As23}] for twisted $1/2$-Cd$_3$As$_2$.
  The color map represents the surface weight~\eqref{eq:Cd3As25} up to two sites.
  The red dashed lines represent the band structure of the unperturbed Hamiltonian~\eqref{eq:Cd3As21} at ${\bm k}_{\perp} = 0$.%
  } \label{fig:Cd3As2_band}
\end{figure}

We diagonalize the total Hamiltonian [Eq.~\eqref{eq:Cd3As21} $+$ Eq.~\eqref{eq:Cd3As23}]
imposing the open boundary conditions in the $x$ and $y$ axes and the periodic one in the $z$ axis.
The numbers of sites in the $x$, $y$, and $z$ axes are $N_x = N_y = 80$ and $N_z = 192$, respectively.
We use $g \theta = 10$, and the corresponding axial magnetic field is $q B_5^z a^2/\hbar = -1.978 \times 10^{-2}$ ($B_5^z = -8.11~{\rm T}$).
This value is reasonable compared with that used in the literature~\cite{PhysRevX.6.041021}.
Although it is difficult to estimate the dimensionless coupling constant $g$, our conclusion is unchanged even if larger or smaller $g \theta$ is used.
Figure~\ref{fig:Cd3As2_band} shows the obtained band structure.
The color map represents the surface weight of the wave function $| u_n(k_z) \rangle$,
\begin{equation}
  \rho^{\rm surf}_n(k_z)
  = \sum_{{\bm x}_{\perp}}^{\rm surf} |\langle {\bm x}_{\perp} | u_n(k_z) \rangle|^2. \label{eq:Cd3As25}
\end{equation}
Here, the summation is taken over the surface up to two sites.
We find the bulk chiral Landau levels and the Fermi arc surface states.

\begin{figure*}
  \centering
  \includegraphics[clip,width=0.98\textwidth]{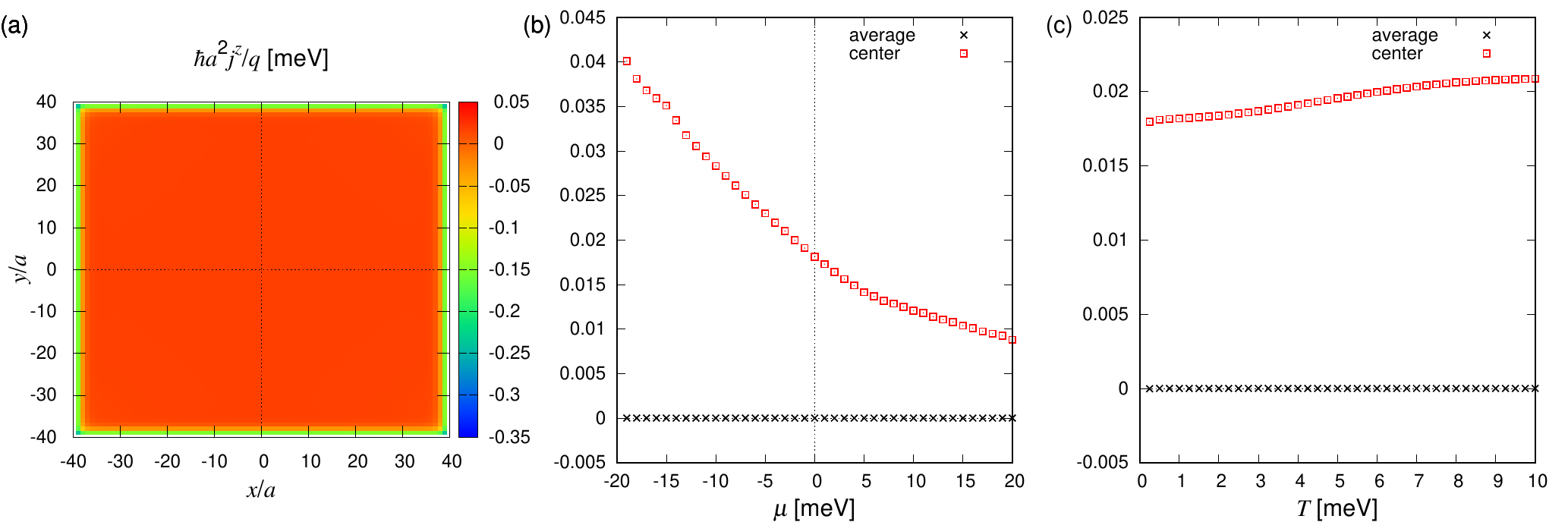}
  \caption{%
  (a) Spatial dependence of the charge current density for twisted $1/2$-Cd$_3$As$_2$ at $\mu = 0~\mev$ and $T = 0.5~\mev$.
  (b) The chemical potential dependence at $T = 0.5~\mev$ and (c) the temperature dependence at $\mu = 0~\mev$
  of the average charge current density (black crosses) and that at the center (red squares).%
  } \label{fig:Cd3As2_cme}
\end{figure*}

\begin{figure*}
  \centering
  \includegraphics[clip,width=0.98\textwidth]{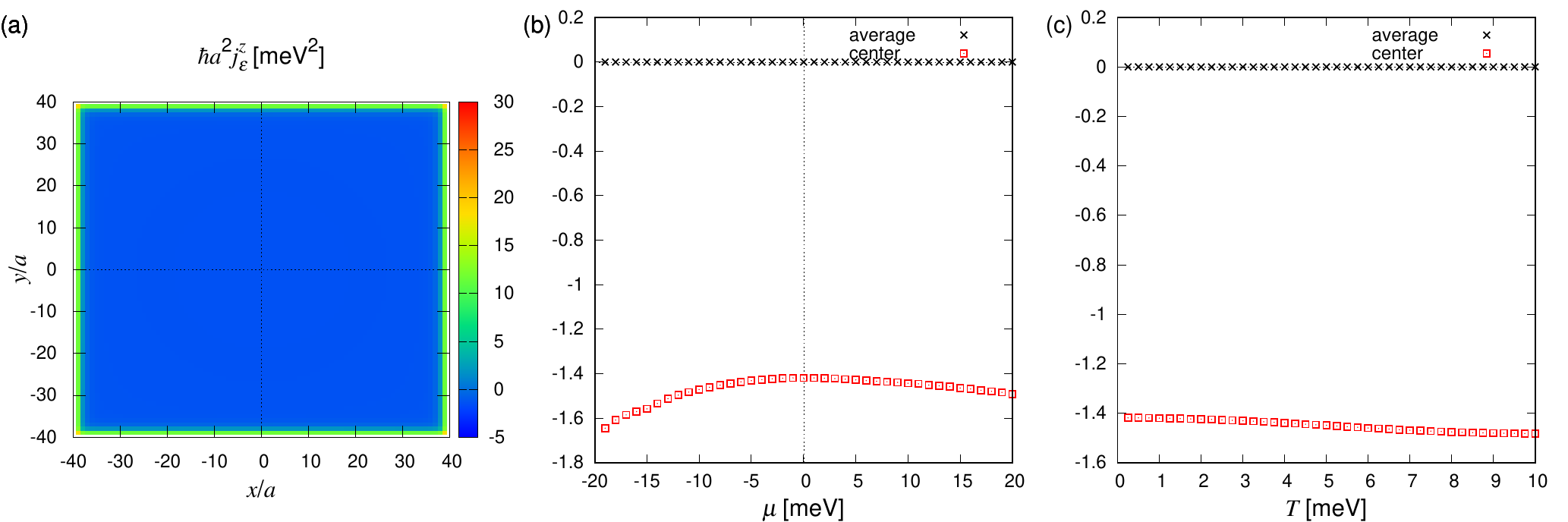}
  \caption{%
  (a) Spatial dependence of the energy current density for twisted $1/2$-Cd$_3$As$_2$ at $\mu = 0~\mev$ and $T = 0.5~\mev$.
  (b) The chemical potential dependence at $T = 0.5~\mev$ and (c) the temperature dependence at $\mu = 0~\mev$
  of the average energy current density (black crosses) and that at the center (red squares).%
  } \label{fig:Cd3As2_ame}
\end{figure*}

The charge and energy current densities are
\begin{subequations} \begin{align}
  j^z({\bm x}_{\perp})
  = & \frac{q}{a^2 c N_z} \sum_{n k_z} v_n^z({\bm x}_{\perp}, k_z) f(\epsilon_n(k_z)), \label{eq:Cd3As26a} \\
  j_{\varepsilon}^z({\bm x}_{\perp})
  = & \frac{1}{a^2 c N_z} \sum_{n k_z} v_n^z({\bm x}_{\perp}, k_z) \epsilon_n(k_z) f(\epsilon_n(k_z)), \label{eq:Cd3As26b}
\end{align} \label{eq:Cd3As26}\end{subequations}
where $v_n^z({\bm x}_{\perp}, k_z) = \hbar^{-1} \langle u_n(k_z) | \partial_{k_z} {\hc H}({\bm x}_{\perp}, k_z) | u_n(k_z) \rangle$,
and $f(\epsilon) = [e^{(\epsilon - \mu)/T} + 1]^{-1}$ is the Fermi distribution function.
The average current densities are obtained by
\begin{equation}
  j_{(\varepsilon) {\rm ave}}^z
  = \frac{1}{N_x N_y} \sum_{{\bm x}_{\perp}} j_{(\varepsilon)}^z({\bm x}_{\perp}). \label{eq:Cd3As27}
\end{equation}
In Fig.~\ref{fig:Cd3As2_cme}(a), the charge current density at the surface is opposite to that inside the bulk.
We show in Figs.~\ref{fig:Cd3As2_cme}(b) and \ref{fig:Cd3As2_cme}(c) that the average charge current density vanishes for any chemical potential $\mu$ and temperature $T$~\cite{PhysRevX.6.041021}.
The same is true for the energy current density as shown in Fig.~\ref{fig:Cd3As2_ame}.
Both the CPME and AME are canceled by the surface contributions and cannot be observed in transport experiments.

\section{Discussion and Summary} \label{sec:discussion}
We attribute the current distributions obtained above to the magnetization currents.
The axial ``gauge'' field in the above two models is a uniquely defined quantity and cannot be identified as a true one~\cite{Landsteiner2016}.
Rather, it is regarded as the (effective) Zeeman field as in Eqs.~\eqref{eq:wilson2} and \eqref{eq:Cd3As23}.
In the presence of a spin-orbit coupling, the orbital magnetization is induced by the Zeeman field as ${\bm M} = \chi_{\rm os} ({\bm A}_5/a)$~\cite{PhysRevLett.97.236805}.
We expect the energy magnetization ${\bm M}_{\varepsilon} = \chi_{\rm es} ({\bm A}_5/a)$ as well.
Now that the Zeeman field is spatially modulated,
the orbital and energy magnetizations are also spatially modulated
and give rise to the charge and energy current densities ${\bm j}_{(\varepsilon)} = {\bm \nabla}_x \times {\bm M}_{(\varepsilon)}$~\cite{Jensen2013}.
However, such magnetization currents cannot be observed in transport experiments.

\begin{figure*}
  \centering
  \includegraphics[clip,width=0.98\textwidth]{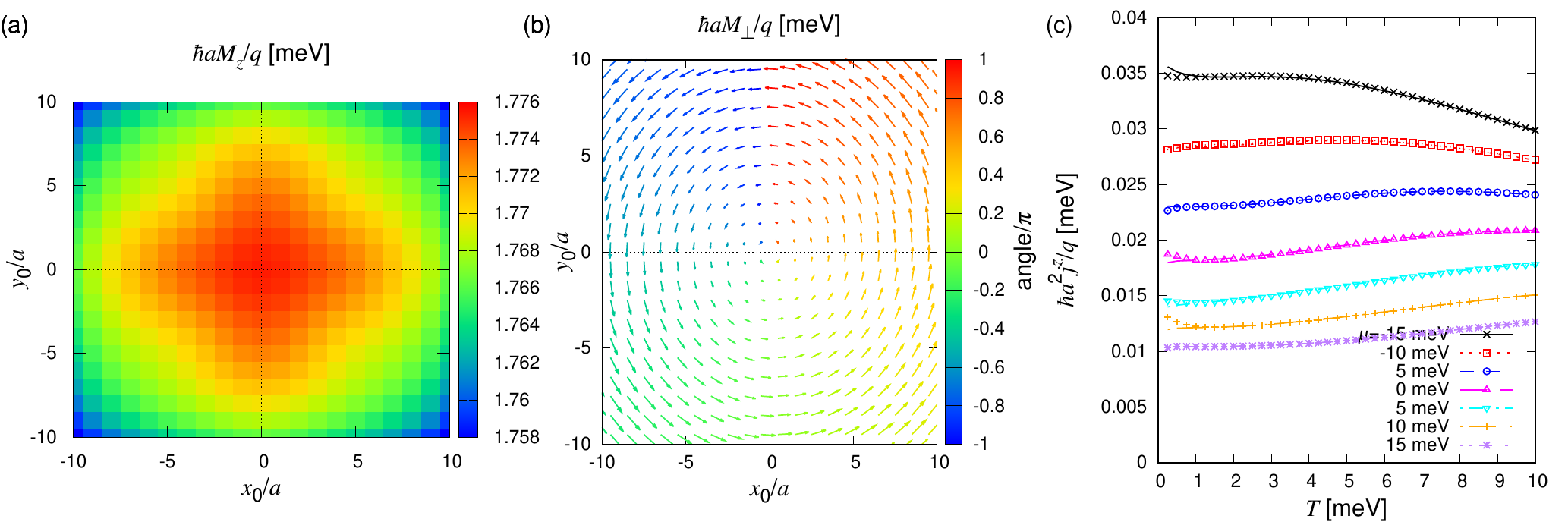}
  \caption{%
  (a) $z$ component and (b) the $x$ and $y$ components of the orbital magnetization for twisted $1/2$-Cd$_3$As$_2$ at $\mu = 0~\mev$ and $T = 0.5~\mev$ as functions of ${\bm x}_{0 \perp}$.
  (c) The temperature dependence of the charge current density at the center.
  The symbols correspond to the magnetization charge current calculated from the orbital magnetization,
  while the lines correspond to that obtained by imposing the open boundary conditions in Sec.~\ref{sec:Cd3As2}.%
  } \label{fig:Cd3As2_chios}
\end{figure*}

\begin{figure*}
  \centering
  \includegraphics[clip,width=0.98\textwidth]{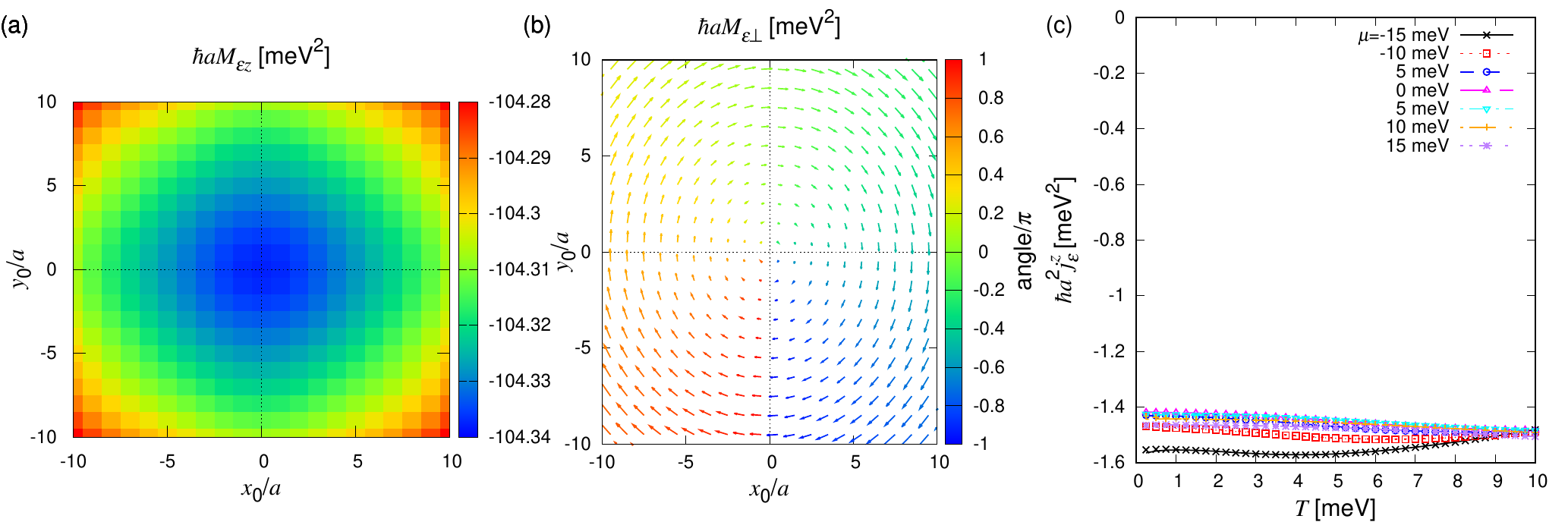}
  \caption{%
  (a) $z$ component and (b) the $x$ and $y$ components of the energy magnetization for twisted $1/2$-Cd$_3$As$_2$ at $\mu = 0~\mev$ and $T = 0.5~\mev$ as functions of ${\bm x}_{0 \perp}$.
  (c) The temperature dependence of the energy current density inside the bulk.
  The symbols correspond to the magnetization energy current calculated from the energy magnetization,
  while the lines correspond to that obtained by imposing the open boundary conditions in Sec.~\ref{sec:Cd3As2}.%
  } \label{fig:Cd3As2_chies}
\end{figure*}

To confirm the above scenario, we calculate the orbital and energy magnetizations for twisted $1/2$-Cd$_3$As$_2$ as follows.
Instead of the total Hamiltonian [Eq.~\eqref{eq:Cd3As21} $+$ Eq.~\eqref{eq:Cd3As23}],
we consider ${\hc H}({\bm k}) = {\hc H}_0({\bm k}) + {\hc H}_1({\bm x}_{0 \perp}, k_z)$ by fixing ${\bm x}_{\perp} = {\bm x}_{0 \perp}$.
${\bm x}_{0 \perp}$ is just the parameter that controls the effective Zeeman field as ${\bm A}_{5 \perp}/a = B_5^z [-y_0, x_0]^{\rm T}/2 a$.
We impose the periodic boundary conditions and calculate the orbital~\cite{PhysRevLett.99.197202}, heat, and energy magnetizations~\cite{Shitade01122014},
\begin{subequations} \begin{align}
  {\bm M}
  = & \frac{q}{\hbar a^2 c N_x N_y N_z} \sum_{n {\bm k}}
  [{\bm m}_n({\bm k}) f(\epsilon_n({\bm k})) \notag \\
  & + {\bm b}_n({\bm k}) f^{(-1)}(\epsilon_n({\bm k}))], \label{eq:chios1a} \\
  {\bm M}_q
  = & \frac{1}{\hbar a^2 c N_x N_y N_z} \sum_{n {\bm k}}
  \{{\bm m}_n({\bm k}) [\epsilon_n({\bm k}) - \mu] f(\epsilon_n({\bm k})) \notag \\
  & + {\bm b}_n({\bm k}) f^{(-2)}(\epsilon_n({\bm k}))\}, \label{eq:chios1b} \\
  {\bm M}_{\varepsilon}
  = & {\bm M}_q + \mu {\bm M}/q. \label{eq:chios1c}
\end{align} \label{eq:chios1}\end{subequations}
Here, the Berry curvature ${\bm b}_n({\bm k})$ and magnetic moment ${\bm m}_n({\bm k})$ are defined as
\begin{subequations} \begin{align}
  {\bm b}_n({\bm k})
  = & i \langle {\bm \nabla}_k u_n({\bm k}) | \times | {\bm \nabla}_k u_n({\bm k}) \rangle, \label{eq:chios2a} \\
  {\bm m}_n({\bm k})
  = & (-i/2) \langle {\bm \nabla}_k u_n({\bm k}) | \notag \\
  & \times [\epsilon_n({\bm k}) - {\hc H}({\bm k})] | {\bm \nabla}_k u_n({\bm k}) \rangle, \label{eq:chios2b}
\end{align} \label{eq:chios2}\end{subequations}
and $f^{(-1)}(\epsilon)$ and $f^{(-2)}(\epsilon)$ are
\begin{subequations} \begin{align}
  f^{(-1)}(\epsilon)
  = & -\int_{\epsilon}^{\infty} d z f(z), \label{eq:chios3a} \\
  f^{(-2)}(\epsilon)
  = & -\int_{\epsilon}^{\infty} d z (z - \mu) f(z). \label{eq:chios3b}
\end{align} \label{eq:chios3}\end{subequations}
The spin-orbit magnetic susceptibility $\chi_{\rm os}$ was already studied in this way~\cite{PhysRevB.93.214434,PhysRevB.99.085205}.
In Fig.~\ref{fig:Cd3As2_chios}(a), the $z$ component of the orbital magnetization is not so affected by the perturbative Zeeman field,
because it mainly comes from the separation of Weyl points~\cite{PhysRevB.93.045201,*PhysRevB.99.209903} determined by $h^z({\bm k})$ in Eq.~\eqref{eq:Cd3As21}.
On the other hand, in Fig.~\ref{fig:Cd3As2_chios}(b), the $x$ and $y$ components are almost proportional to ${\bm A}_{5 \perp}/a$ and show the vortex structure.
We also obtain similar results on the energy magnetization in Figs.~\ref{fig:Cd3As2_chies}(a) and \ref{fig:Cd3As2_chies}(b).
Here, we reinterpret ${\bm x}_{0 \perp}$ as true coordinates
and calculate the discretized versions of $j_{(\varepsilon)}^z = \partial_{x_0} M_{(\varepsilon) y} - \partial_{y_0} M_{(\varepsilon) x}$.
Such an adiabatic approximation is valid only inside the bulk.
In Figs.~\ref{fig:Cd3As2_chios}(c) and \ref{fig:Cd3As2_chies}(c), we show the charge and energy current densities at the center.
The magnetization currents obtained here coincide with those obtained by imposing the open boundary conditions in Sec.~\ref{sec:Cd3As2}.
These results strongly support our scenario of the magnetization currents.

We emphasize that the energy magnetization calculated above is different from the orbital angular momentum studied in Ref.~\cite{PhysRevB.89.081407} in condensed matter systems.
In Ref.~\cite{PhysRevB.89.081407}, it was proposed that in magnetic Weyl semimetals,
the circulating energy current is induced by the azimuthal axial magnetic field at the surface and is observed as the orbital angular momentum.
This proposal relied on the symmetric property of the energy-momentum tensor, i.e., ${\bm j}_{\varepsilon} = {\bm p}$.
However, this property does not hold in condensed matter systems, because the negative-energy states are particles but not antiparticles,
and the energy magnetization and orbital angular momentum are contributed from all the states below the Fermi level, where the approximation of the linear dispersion breaks down.
Indeed, the chemical potential and temperature dependences of the energy current density at the center for twisted $1/2$-Cd$_3$As$_2$ are not described by Eq.~\eqref{eq:cme3},
as shown in Fig.~\ref{fig:Cd3As2_ame}.
In such generic cases, the AME corresponds to the energy magnetization but not directly to the orbital angular momentum.
In contrast to the orbital magnetization, which can be measured by superconducting quantum interference devices, it may be difficult to measure the energy magnetization.

To summarize, we have numerically investigated the CPME and AME using the relativistic Wilson fermion and a lattice model of a Dirac semimetal Cd$_3$As$_2$.
The charge and energy current densities inside the bulk are correctly described by the previous results
[the second term of Eq.~\eqref{eq:cme1b} and Eq.~\eqref{eq:cme3} in the relativistic case].
However, the average current densities completely vanish owing to the surface contributions.
The axial gauge field is regarded as the spatially modulated (effective) Zeeman field and induces the spatially modulated orbital and energy magnetizations.
The CPME and AME currents are the corresponding magnetization currents.
Thus it is impossible to observe the CPME or AME in transport experiments.

What we called anomalous transport phenomena are not transport phenomena in equilibrium.
At the field-theoretical level, Eq.~\eqref{eq:cme1b} is the covariant charge current that is not conserved.
Conserved is the consistent charge current obtained by adding the Bardeen-Zumino (BZ) polynomial~\cite{BARDEEN1984421}.
As a consequence, the CME in the first term of Eq.~\eqref{eq:cme1b} does not occur~\cite{Landsteiner2016,PhysRevLett.118.127601}.
On lattices, we always consider the conserved charge current, and the CME does not occur~\cite{Zhou_2013,PhysRevLett.111.027201}.
The CVE current~\eqref{eq:cme2} is not corrected by the BZ polynomial.
However, it turned out to be the magnetization charge current~\cite{PhysRevB.102.205201}.
The CPME current in the second term of Eq.~\eqref{eq:cme1b} and the AME current~\eqref{eq:cme3} are also the magnetization charge and energy currents.
These three currents do exist but cannot be observed in transport experiments.
%--- Acknowledgments
\begin{acknowledgments}
  We thank M. Chernodub for asking a question that motivated this work when we submitted Ref.~\cite{PhysRevB.102.205201} and T. Hayata for valuable comments on our manuscript.
  This work was supported by the Japan Society for the Promotion of Science KAKENHI (Grant No.~JP18K13508).
  Y.A. is supported by the Leading Initiative for Excellent Young Researchers (LEADER).
\end{acknowledgments}
%--- References
%Control: production of eprint (1) enabled
%

\begin{thebibliography}{61}%
\makeatletter
\providecommand \@ifxundefined [1]{%
 \@ifx{#1\undefined}
}%
\providecommand \@ifnum [1]{%
 \ifnum #1\expandafter \@firstoftwo
 \else \expandafter \@secondoftwo
 \fi
}%
\providecommand \@ifx [1]{%
 \ifx #1\expandafter \@firstoftwo
 \else \expandafter \@secondoftwo
 \fi
}%
\providecommand \natexlab [1]{#1}%
\providecommand \enquote  [1]{``#1''}%
\providecommand \bibnamefont  [1]{#1}%
\providecommand \bibfnamefont [1]{#1}%
\providecommand \citenamefont [1]{#1}%
\providecommand \href@noop [0]{\@secondoftwo}%
\providecommand \href [0]{\begingroup \@sanitize@url \@href}%
\providecommand \@href[1]{\@@startlink{#1}\@@href}%
\providecommand \@@href[1]{\endgroup#1\@@endlink}%
\providecommand \@sanitize@url [0]{\catcode `\\12\catcode `\$12\catcode
  `\&12\catcode `\#12\catcode `\^12\catcode `\_12\catcode `\%12\relax}%
\providecommand \@@startlink[1]{}%
\providecommand \@@endlink[0]{}%
\providecommand \url  [0]{\begingroup\@sanitize@url \@url }%
\providecommand \@url [1]{\endgroup\@href {#1}{\urlprefix }}%
\providecommand \urlprefix  [0]{URL }%
\providecommand \Eprint [0]{\href }%
\providecommand \doibase [0]{https://doi.org/}%
\providecommand \selectlanguage [0]{\@gobble}%
\providecommand \bibinfo  [0]{\@secondoftwo}%
\providecommand \bibfield  [0]{\@secondoftwo}%
\providecommand \translation [1]{[#1]}%
\providecommand \BibitemOpen [0]{}%
\providecommand \bibitemStop [0]{}%
\providecommand \bibitemNoStop [0]{.\EOS\space}%
\providecommand \EOS [0]{\spacefactor3000\relax}%
\providecommand \BibitemShut  [1]{\csname bibitem#1\endcsname}%
\let\auto@bib@innerbib\@empty
%</preamble>
\bibitem [{\citenamefont {Adler}(1969)}]{PhysRev.177.2426}%
  \BibitemOpen
  \bibfield  {author} {\bibinfo {author} {\bibfnamefont {S.~L.}\ \bibnamefont
  {Adler}},\ }\href {https://doi.org/10.1103/PhysRev.177.2426} {\bibfield
  {journal} {\bibinfo  {journal} {Phys. Rev.}\ }\textbf {\bibinfo {volume}
  {177}},\ \bibinfo {pages} {2426} (\bibinfo {year} {1969})}\BibitemShut
  {NoStop}%
\bibitem [{\citenamefont {Bell}\ and\ \citenamefont {Jackiw}(1969)}]{Bell1969}%
  \BibitemOpen
  \bibfield  {author} {\bibinfo {author} {\bibfnamefont {J.~S.}\ \bibnamefont
  {Bell}}\ and\ \bibinfo {author} {\bibfnamefont {R.}~\bibnamefont {Jackiw}},\
  }\href {https://doi.org/10.1007/BF02823296} {\bibfield  {journal} {\bibinfo
  {journal} {Nuovo Cimento A}\ }\textbf {\bibinfo {volume} {60}},\ \bibinfo
  {pages} {47} (\bibinfo {year} {1969})}\BibitemShut {NoStop}%
\bibitem [{\citenamefont {Arsene}\ \emph {et~al.}(2005)\citenamefont {Arsene}
  \emph {et~al.}}]{ARSENE20051}%
  \BibitemOpen
  \bibfield  {author} {\bibinfo {author} {\bibfnamefont {I.}~\bibnamefont
  {Arsene}} \emph {et~al.} (\bibinfo {collaboration} {BRAHMS Collaboration}),\
  }\href {https://doi.org/10.1016/j.nuclphysa.2005.02.130} {\bibfield
  {journal} {\bibinfo  {journal} {Nucl. Phys. A}\ }\textbf {\bibinfo {volume}
  {757}},\ \bibinfo {pages} {1} (\bibinfo {year} {2005})}\BibitemShut {NoStop}%
\bibitem [{\citenamefont {Back}\ \emph {et~al.}(2005)\citenamefont {Back} \emph
  {et~al.}}]{BACK200528}%
  \BibitemOpen
  \bibfield  {author} {\bibinfo {author} {\bibfnamefont {B.~B.}\ \bibnamefont
  {Back}} \emph {et~al.} (\bibinfo {collaboration} {PHOBOS Collaboration}),\
  }\href {https://doi.org/10.1016/j.nuclphysa.2005.03.084} {\bibfield
  {journal} {\bibinfo  {journal} {Nucl. Phys. A}\ }\textbf {\bibinfo {volume}
  {757}},\ \bibinfo {pages} {28} (\bibinfo {year} {2005})}\BibitemShut
  {NoStop}%
\bibitem [{\citenamefont {Adams}\ \emph {et~al.}(2005)\citenamefont {Adams}
  \emph {et~al.}}]{ADAMS2005102}%
  \BibitemOpen
  \bibfield  {author} {\bibinfo {author} {\bibfnamefont {J.}~\bibnamefont
  {Adams}} \emph {et~al.} (\bibinfo {collaboration} {STAR Collaboration}),\
  }\href {https://doi.org/10.1016/j.nuclphysa.2005.03.085} {\bibfield
  {journal} {\bibinfo  {journal} {Nucl. Phys. A}\ }\textbf {\bibinfo {volume}
  {757}},\ \bibinfo {pages} {102} (\bibinfo {year} {2005})}\BibitemShut
  {NoStop}%
\bibitem [{\citenamefont {Adcox}\ \emph {et~al.}(2005)\citenamefont {Adcox}
  \emph {et~al.}}]{ADCOX2005184}%
  \BibitemOpen
  \bibfield  {author} {\bibinfo {author} {\bibfnamefont {K.}~\bibnamefont
  {Adcox}} \emph {et~al.} (\bibinfo {collaboration} {PHENIX Collaboration}),\
  }\href {https://doi.org/10.1016/j.nuclphysa.2005.03.086} {\bibfield
  {journal} {\bibinfo  {journal} {Nucl. Phys. A}\ }\textbf {\bibinfo {volume}
  {757}},\ \bibinfo {pages} {184} (\bibinfo {year} {2005})}\BibitemShut
  {NoStop}%
\bibitem [{\citenamefont {Armitage}\ \emph {et~al.}(2018)\citenamefont
  {Armitage}, \citenamefont {Mele},\ and\ \citenamefont
  {Vishwanath}}]{RevModPhys.90.015001}%
  \BibitemOpen
  \bibfield  {author} {\bibinfo {author} {\bibfnamefont {N.~P.}\ \bibnamefont
  {Armitage}}, \bibinfo {author} {\bibfnamefont {E.~J.}\ \bibnamefont {Mele}},\
  and\ \bibinfo {author} {\bibfnamefont {A.}~\bibnamefont {Vishwanath}},\
  }\href {https://doi.org/10.1103/RevModPhys.90.015001} {\bibfield  {journal}
  {\bibinfo  {journal} {Rev. Mod. Phys.}\ }\textbf {\bibinfo {volume} {90}},\
  \bibinfo {pages} {015001} (\bibinfo {year} {2018})}\BibitemShut {NoStop}%
\bibitem [{\citenamefont {Vilenkin}(1980{\natexlab{a}})}]{PhysRevD.22.3080}%
  \BibitemOpen
  \bibfield  {author} {\bibinfo {author} {\bibfnamefont {A.}~\bibnamefont
  {Vilenkin}},\ }\href {https://doi.org/10.1103/PhysRevD.22.3080} {\bibfield
  {journal} {\bibinfo  {journal} {Phys. Rev. D}\ }\textbf {\bibinfo {volume}
  {22}},\ \bibinfo {pages} {3080} (\bibinfo {year}
  {1980}{\natexlab{a}})}\BibitemShut {NoStop}%
\bibitem [{\citenamefont {Kharzeev}(2006)}]{KHARZEEV2006260}%
  \BibitemOpen
  \bibfield  {author} {\bibinfo {author} {\bibfnamefont {D.}~\bibnamefont
  {Kharzeev}},\ }\href {https://doi.org/10.1016/j.physletb.2005.11.075}
  {\bibfield  {journal} {\bibinfo  {journal} {Phys. Lett. B}\ }\textbf
  {\bibinfo {volume} {633}},\ \bibinfo {pages} {260} (\bibinfo {year}
  {2006})}\BibitemShut {NoStop}%
\bibitem [{\citenamefont {Kharzeev}\ and\ \citenamefont
  {Zhitnitsky}(2007)}]{KHARZEEV200767}%
  \BibitemOpen
  \bibfield  {author} {\bibinfo {author} {\bibfnamefont {D.}~\bibnamefont
  {Kharzeev}}\ and\ \bibinfo {author} {\bibfnamefont {A.}~\bibnamefont
  {Zhitnitsky}},\ }\href {https://doi.org/10.1016/j.nuclphysa.2007.10.001}
  {\bibfield  {journal} {\bibinfo  {journal} {Nucl. Phys. A}\ }\textbf
  {\bibinfo {volume} {797}},\ \bibinfo {pages} {67} (\bibinfo {year}
  {2007})}\BibitemShut {NoStop}%
\bibitem [{\citenamefont {Kharzeev}\ \emph {et~al.}(2008)\citenamefont
  {Kharzeev}, \citenamefont {McLerran},\ and\ \citenamefont
  {Warringa}}]{KHARZEEV2008227}%
  \BibitemOpen
  \bibfield  {author} {\bibinfo {author} {\bibfnamefont {D.~E.}\ \bibnamefont
  {Kharzeev}}, \bibinfo {author} {\bibfnamefont {L.~D.}\ \bibnamefont
  {McLerran}},\ and\ \bibinfo {author} {\bibfnamefont {H.~J.}\ \bibnamefont
  {Warringa}},\ }\href {https://doi.org/10.1016/j.nuclphysa.2008.02.298}
  {\bibfield  {journal} {\bibinfo  {journal} {Nucl. Phys. A}\ }\textbf
  {\bibinfo {volume} {803}},\ \bibinfo {pages} {227} (\bibinfo {year}
  {2008})}\BibitemShut {NoStop}%
\bibitem [{\citenamefont {Fukushima}\ \emph {et~al.}(2008)\citenamefont
  {Fukushima}, \citenamefont {Kharzeev},\ and\ \citenamefont
  {Warringa}}]{PhysRevD.78.074033}%
  \BibitemOpen
  \bibfield  {author} {\bibinfo {author} {\bibfnamefont {K.}~\bibnamefont
  {Fukushima}}, \bibinfo {author} {\bibfnamefont {D.~E.}\ \bibnamefont
  {Kharzeev}},\ and\ \bibinfo {author} {\bibfnamefont {H.~J.}\ \bibnamefont
  {Warringa}},\ }\href {https://doi.org/10.1103/PhysRevD.78.074033} {\bibfield
  {journal} {\bibinfo  {journal} {Phys. Rev. D}\ }\textbf {\bibinfo {volume}
  {78}},\ \bibinfo {pages} {074033} (\bibinfo {year} {2008})}\BibitemShut
  {NoStop}%
\bibitem [{\citenamefont {Pikulin}\ \emph {et~al.}(2016)\citenamefont
  {Pikulin}, \citenamefont {Chen},\ and\ \citenamefont
  {Franz}}]{PhysRevX.6.041021}%
  \BibitemOpen
  \bibfield  {author} {\bibinfo {author} {\bibfnamefont {D.~I.}\ \bibnamefont
  {Pikulin}}, \bibinfo {author} {\bibfnamefont {A.}~\bibnamefont {Chen}},\ and\
  \bibinfo {author} {\bibfnamefont {M.}~\bibnamefont {Franz}},\ }\href
  {https://doi.org/10.1103/PhysRevX.6.041021} {\bibfield  {journal} {\bibinfo
  {journal} {Phys. Rev. X}\ }\textbf {\bibinfo {volume} {6}},\ \bibinfo {pages}
  {041021} (\bibinfo {year} {2016})}\BibitemShut {NoStop}%
\bibitem [{\citenamefont {Grushin}\ \emph {et~al.}(2016)\citenamefont
  {Grushin}, \citenamefont {Venderbos}, \citenamefont {Vishwanath},\ and\
  \citenamefont {Ilan}}]{PhysRevX.6.041046}%
  \BibitemOpen
  \bibfield  {author} {\bibinfo {author} {\bibfnamefont {A.~G.}\ \bibnamefont
  {Grushin}}, \bibinfo {author} {\bibfnamefont {J.~W.~F.}\ \bibnamefont
  {Venderbos}}, \bibinfo {author} {\bibfnamefont {A.}~\bibnamefont
  {Vishwanath}},\ and\ \bibinfo {author} {\bibfnamefont {R.}~\bibnamefont
  {Ilan}},\ }\href {https://doi.org/10.1103/PhysRevX.6.041046} {\bibfield
  {journal} {\bibinfo  {journal} {Phys. Rev. X}\ }\textbf {\bibinfo {volume}
  {6}},\ \bibinfo {pages} {041046} (\bibinfo {year} {2016})}\BibitemShut
  {NoStop}%
\bibitem [{\citenamefont {Zhou}\ \emph {et~al.}(2013)\citenamefont {Zhou},
  \citenamefont {Jiang}, \citenamefont {Niu},\ and\ \citenamefont
  {Shi}}]{Zhou_2013}%
  \BibitemOpen
  \bibfield  {author} {\bibinfo {author} {\bibfnamefont {J.-H.}\ \bibnamefont
  {Zhou}}, \bibinfo {author} {\bibfnamefont {H.}~\bibnamefont {Jiang}},
  \bibinfo {author} {\bibfnamefont {Q.}~\bibnamefont {Niu}},\ and\ \bibinfo
  {author} {\bibfnamefont {J.-R.}\ \bibnamefont {Shi}},\ }\href
  {https://doi.org/10.1088/0256-307x/30/2/027101} {\bibfield  {journal}
  {\bibinfo  {journal} {Chin. Phys. Lett.}\ }\textbf {\bibinfo {volume} {30}},\
  \bibinfo {pages} {027101} (\bibinfo {year} {2013})}\BibitemShut {NoStop}%
\bibitem [{\citenamefont {Vazifeh}\ and\ \citenamefont
  {Franz}(2013)}]{PhysRevLett.111.027201}%
  \BibitemOpen
  \bibfield  {author} {\bibinfo {author} {\bibfnamefont {M.~M.}\ \bibnamefont
  {Vazifeh}}\ and\ \bibinfo {author} {\bibfnamefont {M.}~\bibnamefont
  {Franz}},\ }\href {https://doi.org/10.1103/PhysRevLett.111.027201} {\bibfield
   {journal} {\bibinfo  {journal} {Phys. Rev. Lett.}\ }\textbf {\bibinfo
  {volume} {111}},\ \bibinfo {pages} {027201} (\bibinfo {year}
  {2013})}\BibitemShut {NoStop}%
\bibitem [{\citenamefont {Landsteiner}(2016)}]{Landsteiner2016}%
  \BibitemOpen
  \bibfield  {author} {\bibinfo {author} {\bibfnamefont {K.}~\bibnamefont
  {Landsteiner}},\ }\href {https://doi.org/10.5506/APhysPolB.47.2617}
  {\bibfield  {journal} {\bibinfo  {journal} {Acta Phys. Pol. B}\ }\textbf
  {\bibinfo {volume} {47}},\ \bibinfo {pages} {2617} (\bibinfo {year}
  {2016})}\BibitemShut {NoStop}%
\bibitem [{\citenamefont {Gorbar}\ \emph {et~al.}(2017)\citenamefont {Gorbar},
  \citenamefont {Miransky}, \citenamefont {Shovkovy},\ and\ \citenamefont
  {Sukhachov}}]{PhysRevLett.118.127601}%
  \BibitemOpen
  \bibfield  {author} {\bibinfo {author} {\bibfnamefont {E.~V.}\ \bibnamefont
  {Gorbar}}, \bibinfo {author} {\bibfnamefont {V.~A.}\ \bibnamefont
  {Miransky}}, \bibinfo {author} {\bibfnamefont {I.~A.}\ \bibnamefont
  {Shovkovy}},\ and\ \bibinfo {author} {\bibfnamefont {P.~O.}\ \bibnamefont
  {Sukhachov}},\ }\href {https://doi.org/10.1103/PhysRevLett.118.127601}
  {\bibfield  {journal} {\bibinfo  {journal} {Phys. Rev. Lett.}\ }\textbf
  {\bibinfo {volume} {118}},\ \bibinfo {pages} {127601} (\bibinfo {year}
  {2017})}\BibitemShut {NoStop}%
\bibitem [{\citenamefont {Bohm}(1949)}]{PhysRev.75.502}%
  \BibitemOpen
  \bibfield  {author} {\bibinfo {author} {\bibfnamefont {D.}~\bibnamefont
  {Bohm}},\ }\href {https://doi.org/10.1103/PhysRev.75.502} {\bibfield
  {journal} {\bibinfo  {journal} {Phys. Rev.}\ }\textbf {\bibinfo {volume}
  {75}},\ \bibinfo {pages} {502} (\bibinfo {year} {1949})}\BibitemShut
  {NoStop}%
\bibitem [{\citenamefont {Ohashi}\ and\ \citenamefont
  {Momoi}(1996)}]{JPSJ.65.3254}%
  \BibitemOpen
  \bibfield  {author} {\bibinfo {author} {\bibfnamefont {Y.}~\bibnamefont
  {Ohashi}}\ and\ \bibinfo {author} {\bibfnamefont {T.}~\bibnamefont {Momoi}},\
  }\href {https://doi.org/10.1143/JPSJ.65.3254} {\bibfield  {journal} {\bibinfo
   {journal} {J. Phys. Soc. Jpn.}\ }\textbf {\bibinfo {volume} {65}},\ \bibinfo
  {pages} {3254} (\bibinfo {year} {1996})}\BibitemShut {NoStop}%
\bibitem [{\citenamefont {Yamamoto}(2015)}]{PhysRevD.92.085011}%
  \BibitemOpen
  \bibfield  {author} {\bibinfo {author} {\bibfnamefont {N.}~\bibnamefont
  {Yamamoto}},\ }\href {https://doi.org/10.1103/PhysRevD.92.085011} {\bibfield
  {journal} {\bibinfo  {journal} {Phys. Rev. D}\ }\textbf {\bibinfo {volume}
  {92}},\ \bibinfo {pages} {085011} (\bibinfo {year} {2015})}\BibitemShut
  {NoStop}%
\bibitem [{\citenamefont {Nielsen}\ and\ \citenamefont
  {Ninomiya}(1983)}]{NIELSEN1983389}%
  \BibitemOpen
  \bibfield  {author} {\bibinfo {author} {\bibfnamefont {H.~B.}\ \bibnamefont
  {Nielsen}}\ and\ \bibinfo {author} {\bibfnamefont {M.}~\bibnamefont
  {Ninomiya}},\ }\href {https://doi.org/10.1016/0370-2693(83)91529-0}
  {\bibfield  {journal} {\bibinfo  {journal} {Phys. Lett. B}\ }\textbf
  {\bibinfo {volume} {130}},\ \bibinfo {pages} {389} (\bibinfo {year}
  {1983})}\BibitemShut {NoStop}%
\bibitem [{\citenamefont {Son}\ and\ \citenamefont
  {Spivak}(2013)}]{PhysRevB.88.104412}%
  \BibitemOpen
  \bibfield  {author} {\bibinfo {author} {\bibfnamefont {D.~T.}\ \bibnamefont
  {Son}}\ and\ \bibinfo {author} {\bibfnamefont {B.~Z.}\ \bibnamefont
  {Spivak}},\ }\href {https://doi.org/10.1103/PhysRevB.88.104412} {\bibfield
  {journal} {\bibinfo  {journal} {Phys. Rev. B}\ }\textbf {\bibinfo {volume}
  {88}},\ \bibinfo {pages} {104412} (\bibinfo {year} {2013})}\BibitemShut
  {NoStop}%
\bibitem [{\citenamefont {Huang}\ \emph {et~al.}(2015)\citenamefont {Huang},
  \citenamefont {Zhao}, \citenamefont {Long}, \citenamefont {Wang},
  \citenamefont {Chen}, \citenamefont {Yang}, \citenamefont {Liang},
  \citenamefont {Xue}, \citenamefont {Weng}, \citenamefont {Fang},
  \citenamefont {Dai},\ and\ \citenamefont {Chen}}]{PhysRevx.5.031023}%
  \BibitemOpen
  \bibfield  {author} {\bibinfo {author} {\bibfnamefont {X.}~\bibnamefont
  {Huang}}, \bibinfo {author} {\bibfnamefont {L.}~\bibnamefont {Zhao}},
  \bibinfo {author} {\bibfnamefont {Y.}~\bibnamefont {Long}}, \bibinfo {author}
  {\bibfnamefont {P.}~\bibnamefont {Wang}}, \bibinfo {author} {\bibfnamefont
  {D.}~\bibnamefont {Chen}}, \bibinfo {author} {\bibfnamefont {Z.}~\bibnamefont
  {Yang}}, \bibinfo {author} {\bibfnamefont {H.}~\bibnamefont {Liang}},
  \bibinfo {author} {\bibfnamefont {M.}~\bibnamefont {Xue}}, \bibinfo {author}
  {\bibfnamefont {H.}~\bibnamefont {Weng}}, \bibinfo {author} {\bibfnamefont
  {Z.}~\bibnamefont {Fang}}, \bibinfo {author} {\bibfnamefont {X.}~\bibnamefont
  {Dai}},\ and\ \bibinfo {author} {\bibfnamefont {G.}~\bibnamefont {Chen}},\
  }\href {https://doi.org/10.1103/PhysRevX.5.031023} {\bibfield  {journal}
  {\bibinfo  {journal} {Phys. Rev. X}\ }\textbf {\bibinfo {volume} {5}},\
  \bibinfo {pages} {031023} (\bibinfo {year} {2015})}\BibitemShut {NoStop}%
\bibitem [{\citenamefont {Du}\ \emph {et~al.}(2016)\citenamefont {Du},
  \citenamefont {Wang}, \citenamefont {Chen}, \citenamefont {Mao},
  \citenamefont {Khan}, \citenamefont {Xu}, \citenamefont {Zhou}, \citenamefont
  {Zhang}, \citenamefont {Yang}, \citenamefont {Chen}, \citenamefont {Feng},\
  and\ \citenamefont {Fang}}]{Du2016}%
  \BibitemOpen
  \bibfield  {author} {\bibinfo {author} {\bibfnamefont {J.}~\bibnamefont
  {Du}}, \bibinfo {author} {\bibfnamefont {H.}~\bibnamefont {Wang}}, \bibinfo
  {author} {\bibfnamefont {Q.}~\bibnamefont {Chen}}, \bibinfo {author}
  {\bibfnamefont {Q.}~\bibnamefont {Mao}}, \bibinfo {author} {\bibfnamefont
  {R.}~\bibnamefont {Khan}}, \bibinfo {author} {\bibfnamefont {B.}~\bibnamefont
  {Xu}}, \bibinfo {author} {\bibfnamefont {Y.}~\bibnamefont {Zhou}}, \bibinfo
  {author} {\bibfnamefont {Y.}~\bibnamefont {Zhang}}, \bibinfo {author}
  {\bibfnamefont {J.}~\bibnamefont {Yang}}, \bibinfo {author} {\bibfnamefont
  {B.}~\bibnamefont {Chen}}, \bibinfo {author} {\bibfnamefont {C.}~\bibnamefont
  {Feng}},\ and\ \bibinfo {author} {\bibfnamefont {M.}~\bibnamefont {Fang}},\
  }\href {https://doi.org/10.1007/s11433-016-5798-4} {\bibfield  {journal}
  {\bibinfo  {journal} {Sci. China Phys. Mech. Astron.}\ }\textbf {\bibinfo
  {volume} {59}},\ \bibinfo {pages} {657406} (\bibinfo {year}
  {2016})}\BibitemShut {NoStop}%
\bibitem [{\citenamefont {Zhang}\ \emph {et~al.}(2016)\citenamefont {Zhang},
  \citenamefont {Xu}, \citenamefont {Belopolski}, \citenamefont {Yuan},
  \citenamefont {Lin}, \citenamefont {Tong}, \citenamefont {Bian},
  \citenamefont {Alidoust}, \citenamefont {Lee}, \citenamefont {Huang},
  \citenamefont {Chang}, \citenamefont {Chang}, \citenamefont {Hsu},
  \citenamefont {Jeng}, \citenamefont {Neupane}, \citenamefont {Sanchez},
  \citenamefont {Zheng}, \citenamefont {Wang}, \citenamefont {Lin},
  \citenamefont {Zhang}, \citenamefont {Lu}, \citenamefont {Shen},
  \citenamefont {Neupert}, \citenamefont {Hasan},\ and\ \citenamefont
  {Jia}}]{Zhang2016}%
  \BibitemOpen
  \bibfield  {author} {\bibinfo {author} {\bibfnamefont {C.-L.}\ \bibnamefont
  {Zhang}}, \bibinfo {author} {\bibfnamefont {S.-Y.}\ \bibnamefont {Xu}},
  \bibinfo {author} {\bibfnamefont {I.}~\bibnamefont {Belopolski}}, \bibinfo
  {author} {\bibfnamefont {Z.}~\bibnamefont {Yuan}}, \bibinfo {author}
  {\bibfnamefont {Z.}~\bibnamefont {Lin}}, \bibinfo {author} {\bibfnamefont
  {B.}~\bibnamefont {Tong}}, \bibinfo {author} {\bibfnamefont {G.}~\bibnamefont
  {Bian}}, \bibinfo {author} {\bibfnamefont {N.}~\bibnamefont {Alidoust}},
  \bibinfo {author} {\bibfnamefont {C.-C.}\ \bibnamefont {Lee}}, \bibinfo
  {author} {\bibfnamefont {S.-M.}\ \bibnamefont {Huang}}, \bibinfo {author}
  {\bibfnamefont {T.-R.}\ \bibnamefont {Chang}}, \bibinfo {author}
  {\bibfnamefont {G.}~\bibnamefont {Chang}}, \bibinfo {author} {\bibfnamefont
  {C.-H.}\ \bibnamefont {Hsu}}, \bibinfo {author} {\bibfnamefont {H.-T.}\
  \bibnamefont {Jeng}}, \bibinfo {author} {\bibfnamefont {M.}~\bibnamefont
  {Neupane}}, \bibinfo {author} {\bibfnamefont {D.~S.}\ \bibnamefont
  {Sanchez}}, \bibinfo {author} {\bibfnamefont {H.}~\bibnamefont {Zheng}},
  \bibinfo {author} {\bibfnamefont {J.}~\bibnamefont {Wang}}, \bibinfo {author}
  {\bibfnamefont {H.}~\bibnamefont {Lin}}, \bibinfo {author} {\bibfnamefont
  {C.}~\bibnamefont {Zhang}}, \bibinfo {author} {\bibfnamefont {H.-Z.}\
  \bibnamefont {Lu}}, \bibinfo {author} {\bibfnamefont {S.-Q.}\ \bibnamefont
  {Shen}}, \bibinfo {author} {\bibfnamefont {T.}~\bibnamefont {Neupert}},
  \bibinfo {author} {\bibfnamefont {M.~Z.}\ \bibnamefont {Hasan}},\ and\
  \bibinfo {author} {\bibfnamefont {S.}~\bibnamefont {Jia}},\ }\href
  {https://doi.org/10.1038/ncomms10735} {\bibfield  {journal} {\bibinfo
  {journal} {Nat. Commun.}\ }\textbf {\bibinfo {volume} {7}},\ \bibinfo {pages}
  {10735} (\bibinfo {year} {2016})}\BibitemShut {NoStop}%
\bibitem [{\citenamefont {Wang}\ \emph {et~al.}(2016)\citenamefont {Wang},
  \citenamefont {Zheng}, \citenamefont {Shen}, \citenamefont {Lu},
  \citenamefont {Fang}, \citenamefont {Sheng}, \citenamefont {Zhou},
  \citenamefont {Yang}, \citenamefont {Li}, \citenamefont {Feng},\ and\
  \citenamefont {Xu}}]{PhysRevB.93.121112}%
  \BibitemOpen
  \bibfield  {author} {\bibinfo {author} {\bibfnamefont {Z.}~\bibnamefont
  {Wang}}, \bibinfo {author} {\bibfnamefont {Y.}~\bibnamefont {Zheng}},
  \bibinfo {author} {\bibfnamefont {Z.}~\bibnamefont {Shen}}, \bibinfo {author}
  {\bibfnamefont {Y.}~\bibnamefont {Lu}}, \bibinfo {author} {\bibfnamefont
  {H.}~\bibnamefont {Fang}}, \bibinfo {author} {\bibfnamefont {F.}~\bibnamefont
  {Sheng}}, \bibinfo {author} {\bibfnamefont {Y.}~\bibnamefont {Zhou}},
  \bibinfo {author} {\bibfnamefont {X.}~\bibnamefont {Yang}}, \bibinfo {author}
  {\bibfnamefont {Y.}~\bibnamefont {Li}}, \bibinfo {author} {\bibfnamefont
  {C.}~\bibnamefont {Feng}},\ and\ \bibinfo {author} {\bibfnamefont {Z.-A.}\
  \bibnamefont {Xu}},\ }\href {https://doi.org/10.1103/PhysRevB.93.121112}
  {\bibfield  {journal} {\bibinfo  {journal} {Phys. Rev. B}\ }\textbf {\bibinfo
  {volume} {93}},\ \bibinfo {pages} {121112(R)} (\bibinfo {year}
  {2016})}\BibitemShut {NoStop}%
\bibitem [{\citenamefont {Vilenkin}(1979)}]{PhysRevD.20.1807}%
  \BibitemOpen
  \bibfield  {author} {\bibinfo {author} {\bibfnamefont {A.}~\bibnamefont
  {Vilenkin}},\ }\href {https://doi.org/10.1103/PhysRevD.20.1807} {\bibfield
  {journal} {\bibinfo  {journal} {Phys. Rev. D}\ }\textbf {\bibinfo {volume}
  {20}},\ \bibinfo {pages} {1807} (\bibinfo {year} {1979})}\BibitemShut
  {NoStop}%
\bibitem [{\citenamefont {Vilenkin}(1980{\natexlab{b}})}]{PhysRevD.21.2260}%
  \BibitemOpen
  \bibfield  {author} {\bibinfo {author} {\bibfnamefont {A.}~\bibnamefont
  {Vilenkin}},\ }\href {https://doi.org/10.1103/PhysRevD.21.2260} {\bibfield
  {journal} {\bibinfo  {journal} {Phys. Rev. D}\ }\textbf {\bibinfo {volume}
  {21}},\ \bibinfo {pages} {2260} (\bibinfo {year}
  {1980}{\natexlab{b}})}\BibitemShut {NoStop}%
\bibitem [{\citenamefont {Erdmenger}\ \emph {et~al.}(2009)\citenamefont
  {Erdmenger}, \citenamefont {Haack}, \citenamefont {Kaminski},\ and\
  \citenamefont {Yarom}}]{Erdmenger2009}%
  \BibitemOpen
  \bibfield  {author} {\bibinfo {author} {\bibfnamefont {J.}~\bibnamefont
  {Erdmenger}}, \bibinfo {author} {\bibfnamefont {M.}~\bibnamefont {Haack}},
  \bibinfo {author} {\bibfnamefont {M.}~\bibnamefont {Kaminski}},\ and\
  \bibinfo {author} {\bibfnamefont {A.}~\bibnamefont {Yarom}},\ }\href
  {https://doi.org/10.1088/1126-6708/2009/01/055} {\bibfield  {journal}
  {\bibinfo  {journal} {J. High Energy Phys.}\ }\textbf {\bibinfo {volume}
  {2009}},\ \bibinfo {pages} {055}}\BibitemShut {NoStop}%
\bibitem [{\citenamefont {Banerjee}\ \emph {et~al.}(2011)\citenamefont
  {Banerjee}, \citenamefont {Bhattacharya}, \citenamefont {Bhattacharyya},
  \citenamefont {Dutta}, \citenamefont {Loganayagam},\ and\ \citenamefont
  {Sur\'owka}}]{Banerjee2011}%
  \BibitemOpen
  \bibfield  {author} {\bibinfo {author} {\bibfnamefont {N.}~\bibnamefont
  {Banerjee}}, \bibinfo {author} {\bibfnamefont {J.}~\bibnamefont
  {Bhattacharya}}, \bibinfo {author} {\bibfnamefont {S.}~\bibnamefont
  {Bhattacharyya}}, \bibinfo {author} {\bibfnamefont {S.}~\bibnamefont
  {Dutta}}, \bibinfo {author} {\bibfnamefont {R.}~\bibnamefont {Loganayagam}},\
  and\ \bibinfo {author} {\bibfnamefont {P.}~\bibnamefont {Sur\'owka}},\ }\href
  {https://doi.org/10.1007/JHEP01(2011)094} {\bibfield  {journal} {\bibinfo
  {journal} {J. High Energy Phys.}\ }\textbf {\bibinfo {volume} {2011}},\
  \bibinfo {pages} {94}}\BibitemShut {NoStop}%
\bibitem [{\citenamefont {Son}\ and\ \citenamefont
  {Sur\'owka}(2009)}]{PhysRevLett.103.191601}%
  \BibitemOpen
  \bibfield  {author} {\bibinfo {author} {\bibfnamefont {D.~T.}\ \bibnamefont
  {Son}}\ and\ \bibinfo {author} {\bibfnamefont {P.}~\bibnamefont
  {Sur\'owka}},\ }\href {https://doi.org/10.1103/PhysRevLett.103.191601}
  {\bibfield  {journal} {\bibinfo  {journal} {Phys. Rev. Lett.}\ }\textbf
  {\bibinfo {volume} {103}},\ \bibinfo {pages} {191601} (\bibinfo {year}
  {2009})}\BibitemShut {NoStop}%
\bibitem [{\citenamefont {Landsteiner}\ \emph
  {et~al.}(2011{\natexlab{a}})\citenamefont {Landsteiner}, \citenamefont
  {Meg\'{\i}as},\ and\ \citenamefont {Pena-Benitez}}]{PhysRevLett.107.021601}%
  \BibitemOpen
  \bibfield  {author} {\bibinfo {author} {\bibfnamefont {K.}~\bibnamefont
  {Landsteiner}}, \bibinfo {author} {\bibfnamefont {E.}~\bibnamefont
  {Meg\'{\i}as}},\ and\ \bibinfo {author} {\bibfnamefont {F.}~\bibnamefont
  {Pena-Benitez}},\ }\href {https://doi.org/10.1103/PhysRevLett.107.021601}
  {\bibfield  {journal} {\bibinfo  {journal} {Phys. Rev. Lett.}\ }\textbf
  {\bibinfo {volume} {107}},\ \bibinfo {pages} {021601} (\bibinfo {year}
  {2011}{\natexlab{a}})}\BibitemShut {NoStop}%
\bibitem [{\citenamefont {Landsteiner}\ \emph
  {et~al.}(2011{\natexlab{b}})\citenamefont {Landsteiner}, \citenamefont
  {Eugenio~Meg\'{\i}as},\ and\ \citenamefont {Pena-Benitez}}]{Landsteiner2011}%
  \BibitemOpen
  \bibfield  {author} {\bibinfo {author} {\bibfnamefont {K.}~\bibnamefont
  {Landsteiner}}, \bibinfo {author} {\bibfnamefont {L.~M.}\ \bibnamefont
  {Eugenio~Meg\'{\i}as}},\ and\ \bibinfo {author} {\bibfnamefont
  {F.}~\bibnamefont {Pena-Benitez}},\ }\href
  {https://doi.org/10.1007/JHEP09(2011)121} {\bibfield  {journal} {\bibinfo
  {journal} {J. High Energy Phys.}\ }\textbf {\bibinfo {volume} {2011}},\
  \bibinfo {pages} {121}}\BibitemShut {NoStop}%
\bibitem [{\citenamefont {Chen}\ \emph {et~al.}(2014)\citenamefont {Chen},
  \citenamefont {Son}, \citenamefont {Stephanov}, \citenamefont {Yee},\ and\
  \citenamefont {Yin}}]{PhysRevLett.113.182302}%
  \BibitemOpen
  \bibfield  {author} {\bibinfo {author} {\bibfnamefont {J.-Y.}\ \bibnamefont
  {Chen}}, \bibinfo {author} {\bibfnamefont {D.~T.}\ \bibnamefont {Son}},
  \bibinfo {author} {\bibfnamefont {M.~A.}\ \bibnamefont {Stephanov}}, \bibinfo
  {author} {\bibfnamefont {H.-U.}\ \bibnamefont {Yee}},\ and\ \bibinfo {author}
  {\bibfnamefont {Y.}~\bibnamefont {Yin}},\ }\href
  {https://doi.org/10.1103/PhysRevLett.113.182302} {\bibfield  {journal}
  {\bibinfo  {journal} {Phys. Rev. Lett.}\ }\textbf {\bibinfo {volume} {113}},\
  \bibinfo {pages} {182302} (\bibinfo {year} {2014})}\BibitemShut {NoStop}%
\bibitem [{\citenamefont {Shitade}\ \emph {et~al.}(2020)\citenamefont
  {Shitade}, \citenamefont {Mameda},\ and\ \citenamefont
  {Hayata}}]{PhysRevB.102.205201}%
  \BibitemOpen
  \bibfield  {author} {\bibinfo {author} {\bibfnamefont {A.}~\bibnamefont
  {Shitade}}, \bibinfo {author} {\bibfnamefont {K.}~\bibnamefont {Mameda}},\
  and\ \bibinfo {author} {\bibfnamefont {T.}~\bibnamefont {Hayata}},\ }\href
  {https://doi.org/10.1103/PhysRevB.102.205201} {\bibfield  {journal} {\bibinfo
   {journal} {Phys. Rev. B}\ }\textbf {\bibinfo {volume} {102}},\ \bibinfo
  {pages} {205201} (\bibinfo {year} {2020})}\BibitemShut {NoStop}%
\bibitem [{\citenamefont {Braguta}\ \emph {et~al.}(2013)\citenamefont
  {Braguta}, \citenamefont {Chernodub}, \citenamefont {Landsteiner},
  \citenamefont {Polikarpov},\ and\ \citenamefont
  {Ulybyshev}}]{PhysRevD.88.071501}%
  \BibitemOpen
  \bibfield  {author} {\bibinfo {author} {\bibfnamefont {V.}~\bibnamefont
  {Braguta}}, \bibinfo {author} {\bibfnamefont {M.~N.}\ \bibnamefont
  {Chernodub}}, \bibinfo {author} {\bibfnamefont {K.}~\bibnamefont
  {Landsteiner}}, \bibinfo {author} {\bibfnamefont {M.~I.}\ \bibnamefont
  {Polikarpov}},\ and\ \bibinfo {author} {\bibfnamefont {M.~V.}\ \bibnamefont
  {Ulybyshev}},\ }\href {https://doi.org/10.1103/PhysRevD.88.071501} {\bibfield
   {journal} {\bibinfo  {journal} {Phys. Rev. D}\ }\textbf {\bibinfo {volume}
  {88}},\ \bibinfo {pages} {071501(R)} (\bibinfo {year} {2013})}\BibitemShut
  {NoStop}%
\bibitem [{\citenamefont {Buividovich}(2015)}]{Buividovich_2015}%
  \BibitemOpen
  \bibfield  {author} {\bibinfo {author} {\bibfnamefont {P.~V.}\ \bibnamefont
  {Buividovich}},\ }\href {https://doi.org/10.1088/1742-6596/607/1/012018}
  {\bibfield  {journal} {\bibinfo  {journal} {J. Phys. Conf. Ser.}\ }\textbf
  {\bibinfo {volume} {607}},\ \bibinfo {pages} {012018} (\bibinfo {year}
  {2015})}\BibitemShut {NoStop}%
\bibitem [{\citenamefont {Chernodub}\ \emph {et~al.}(2014)\citenamefont
  {Chernodub}, \citenamefont {Cortijo}, \citenamefont {Grushin}, \citenamefont
  {Landsteiner},\ and\ \citenamefont {Vozmediano}}]{PhysRevB.89.081407}%
  \BibitemOpen
  \bibfield  {author} {\bibinfo {author} {\bibfnamefont {M.~N.}\ \bibnamefont
  {Chernodub}}, \bibinfo {author} {\bibfnamefont {A.}~\bibnamefont {Cortijo}},
  \bibinfo {author} {\bibfnamefont {A.~G.}\ \bibnamefont {Grushin}}, \bibinfo
  {author} {\bibfnamefont {K.}~\bibnamefont {Landsteiner}},\ and\ \bibinfo
  {author} {\bibfnamefont {M.~A.~H.}\ \bibnamefont {Vozmediano}},\ }\href
  {https://doi.org/10.1103/PhysRevB.89.081407} {\bibfield  {journal} {\bibinfo
  {journal} {Phys. Rev. B}\ }\textbf {\bibinfo {volume} {89}},\ \bibinfo
  {pages} {081407(R)} (\bibinfo {year} {2014})}\BibitemShut {NoStop}%
\bibitem [{\citenamefont {Liu}\ \emph {et~al.}(2013)\citenamefont {Liu},
  \citenamefont {Ye},\ and\ \citenamefont {Qi}}]{PhysRevB.87.235306}%
  \BibitemOpen
  \bibfield  {author} {\bibinfo {author} {\bibfnamefont {C.-X.}\ \bibnamefont
  {Liu}}, \bibinfo {author} {\bibfnamefont {P.}~\bibnamefont {Ye}},\ and\
  \bibinfo {author} {\bibfnamefont {X.-L.}\ \bibnamefont {Qi}},\ }\href
  {https://doi.org/10.1103/PhysRevB.87.235306} {\bibfield  {journal} {\bibinfo
  {journal} {Phys. Rev. B}\ }\textbf {\bibinfo {volume} {87}},\ \bibinfo
  {pages} {235306} (\bibinfo {year} {2013})}\BibitemShut {NoStop}%
\bibitem [{\citenamefont {Liu}\ \emph {et~al.}(2015)\citenamefont {Liu},
  \citenamefont {Ye},\ and\ \citenamefont {Qi}}]{PhysRevB.92.119904}%
  \BibitemOpen
  \bibfield  {author} {\bibinfo {author} {\bibfnamefont {C.-X.}\ \bibnamefont
  {Liu}}, \bibinfo {author} {\bibfnamefont {P.}~\bibnamefont {Ye}},\ and\
  \bibinfo {author} {\bibfnamefont {X.-L.}\ \bibnamefont {Qi}},\ }\href
  {https://doi.org/10.1103/PhysRevB.92.119904} {\bibfield  {journal} {\bibinfo
  {journal} {Phys. Rev. B}\ }\textbf {\bibinfo {volume} {92}},\ \bibinfo
  {pages} {119904(E)} (\bibinfo {year} {2015})}\BibitemShut {NoStop}%
\bibitem [{\citenamefont {Cortijo}\ \emph {et~al.}(2015)\citenamefont
  {Cortijo}, \citenamefont {Ferreir\'os}, \citenamefont {Landsteiner},\ and\
  \citenamefont {Vozmediano}}]{PhysRevLett.115.177202}%
  \BibitemOpen
  \bibfield  {author} {\bibinfo {author} {\bibfnamefont {A.}~\bibnamefont
  {Cortijo}}, \bibinfo {author} {\bibfnamefont {Y.}~\bibnamefont
  {Ferreir\'os}}, \bibinfo {author} {\bibfnamefont {K.}~\bibnamefont
  {Landsteiner}},\ and\ \bibinfo {author} {\bibfnamefont {M.~A.~H.}\
  \bibnamefont {Vozmediano}},\ }\href
  {https://doi.org/10.1103/PhysRevLett.115.177202} {\bibfield  {journal}
  {\bibinfo  {journal} {Phys. Rev. Lett.}\ }\textbf {\bibinfo {volume} {115}},\
  \bibinfo {pages} {177202} (\bibinfo {year} {2015})}\BibitemShut {NoStop}%
\bibitem [{\citenamefont {Araki}\ \emph {et~al.}(2016)\citenamefont {Araki},
  \citenamefont {Yoshida},\ and\ \citenamefont {Nomura}}]{PhysRevB.94.115312}%
  \BibitemOpen
  \bibfield  {author} {\bibinfo {author} {\bibfnamefont {Y.}~\bibnamefont
  {Araki}}, \bibinfo {author} {\bibfnamefont {A.}~\bibnamefont {Yoshida}},\
  and\ \bibinfo {author} {\bibfnamefont {K.}~\bibnamefont {Nomura}},\ }\href
  {https://doi.org/10.1103/PhysRevB.94.115312} {\bibfield  {journal} {\bibinfo
  {journal} {Phys. Rev. B}\ }\textbf {\bibinfo {volume} {94}},\ \bibinfo
  {pages} {115312} (\bibinfo {year} {2016})}\BibitemShut {NoStop}%
\bibitem [{\citenamefont {Ilan}\ \emph {et~al.}(2020)\citenamefont {Ilan},
  \citenamefont {Grushin},\ and\ \citenamefont {Pikulin}}]{Ilan2019}%
  \BibitemOpen
  \bibfield  {author} {\bibinfo {author} {\bibfnamefont {R.}~\bibnamefont
  {Ilan}}, \bibinfo {author} {\bibfnamefont {A.~G.}\ \bibnamefont {Grushin}},\
  and\ \bibinfo {author} {\bibfnamefont {D.~I.}\ \bibnamefont {Pikulin}},\
  }\href {https://doi.org/10.1038/s42254-019-0121-8} {\bibfield  {journal}
  {\bibinfo  {journal} {Nat. Rev. Phys.}\ }\textbf {\bibinfo {volume} {2}},\
  \bibinfo {pages} {29} (\bibinfo {year} {2020})}\BibitemShut {NoStop}%
\bibitem [{\citenamefont {Wilson}(1974)}]{PhysRevD.10.2445}%
  \BibitemOpen
  \bibfield  {author} {\bibinfo {author} {\bibfnamefont {K.~G.}\ \bibnamefont
  {Wilson}},\ }\href {https://doi.org/10.1103/PhysRevD.10.2445} {\bibfield
  {journal} {\bibinfo  {journal} {Phys. Rev. D}\ }\textbf {\bibinfo {volume}
  {10}},\ \bibinfo {pages} {2445} (\bibinfo {year} {1974})}\BibitemShut
  {NoStop}%
\bibitem [{\citenamefont {Neuberger}(1998)}]{NEUBERGER1998141}%
  \BibitemOpen
  \bibfield  {author} {\bibinfo {author} {\bibfnamefont {H.}~\bibnamefont
  {Neuberger}},\ }\href {https://doi.org/10.1016/S0370-2693(97)01368-3}
  {\bibfield  {journal} {\bibinfo  {journal} {Phys. Lett. B}\ }\textbf
  {\bibinfo {volume} {417}},\ \bibinfo {pages} {141} (\bibinfo {year}
  {1998})}\BibitemShut {NoStop}%
\bibitem [{\citenamefont {Wang}\ \emph {et~al.}(2013)\citenamefont {Wang},
  \citenamefont {Weng}, \citenamefont {Wu}, \citenamefont {Dai},\ and\
  \citenamefont {Fang}}]{PhysRevB.88.125427}%
  \BibitemOpen
  \bibfield  {author} {\bibinfo {author} {\bibfnamefont {Z.}~\bibnamefont
  {Wang}}, \bibinfo {author} {\bibfnamefont {H.}~\bibnamefont {Weng}}, \bibinfo
  {author} {\bibfnamefont {Q.}~\bibnamefont {Wu}}, \bibinfo {author}
  {\bibfnamefont {X.}~\bibnamefont {Dai}},\ and\ \bibinfo {author}
  {\bibfnamefont {Z.}~\bibnamefont {Fang}},\ }\href
  {https://doi.org/10.1103/PhysRevB.88.125427} {\bibfield  {journal} {\bibinfo
  {journal} {Phys. Rev. B}\ }\textbf {\bibinfo {volume} {88}},\ \bibinfo
  {pages} {125427} (\bibinfo {year} {2013})}\BibitemShut {NoStop}%
\bibitem [{\citenamefont {Neupane}\ \emph {et~al.}(2014)\citenamefont
  {Neupane}, \citenamefont {Xu}, \citenamefont {Sankar}, \citenamefont
  {Alidoust}, \citenamefont {Bian}, \citenamefont {Liu}, \citenamefont
  {Belopolski}, \citenamefont {Chang}, \citenamefont {Jeng}, \citenamefont
  {Lin}, \citenamefont {Bansil}, \citenamefont {Chou},\ and\ \citenamefont
  {Hasan}}]{Neupane2014}%
  \BibitemOpen
  \bibfield  {author} {\bibinfo {author} {\bibfnamefont {M.}~\bibnamefont
  {Neupane}}, \bibinfo {author} {\bibfnamefont {S.-Y.}\ \bibnamefont {Xu}},
  \bibinfo {author} {\bibfnamefont {R.}~\bibnamefont {Sankar}}, \bibinfo
  {author} {\bibfnamefont {N.}~\bibnamefont {Alidoust}}, \bibinfo {author}
  {\bibfnamefont {G.}~\bibnamefont {Bian}}, \bibinfo {author} {\bibfnamefont
  {C.}~\bibnamefont {Liu}}, \bibinfo {author} {\bibfnamefont {I.}~\bibnamefont
  {Belopolski}}, \bibinfo {author} {\bibfnamefont {T.-R.}\ \bibnamefont
  {Chang}}, \bibinfo {author} {\bibfnamefont {H.-T.}\ \bibnamefont {Jeng}},
  \bibinfo {author} {\bibfnamefont {H.}~\bibnamefont {Lin}}, \bibinfo {author}
  {\bibfnamefont {A.}~\bibnamefont {Bansil}}, \bibinfo {author} {\bibfnamefont
  {F.}~\bibnamefont {Chou}},\ and\ \bibinfo {author} {\bibfnamefont {M.~Z.}\
  \bibnamefont {Hasan}},\ }\href {https://doi.org/10.1038/ncomms4786}
  {\bibfield  {journal} {\bibinfo  {journal} {Nat. Commun.}\ }\textbf {\bibinfo
  {volume} {5}},\ \bibinfo {pages} {3786} (\bibinfo {year} {2014})}\BibitemShut
  {NoStop}%
\bibitem [{\citenamefont {Liu}\ \emph {et~al.}(2014)\citenamefont {Liu},
  \citenamefont {Jiang}, \citenamefont {Zhou}, \citenamefont {Wang},
  \citenamefont {Zhang}, \citenamefont {Weng}, \citenamefont {Prabhakaran},
  \citenamefont {Mo}, \citenamefont {Peng}, \citenamefont {Dudin},
  \citenamefont {Kim}, \citenamefont {Hoesch}, \citenamefont {Fang},
  \citenamefont {Dai}, \citenamefont {Shen}, \citenamefont {Feng},
  \citenamefont {Hussain},\ and\ \citenamefont {Chen}}]{Liu2014}%
  \BibitemOpen
  \bibfield  {author} {\bibinfo {author} {\bibfnamefont {Z.~K.}\ \bibnamefont
  {Liu}}, \bibinfo {author} {\bibfnamefont {J.}~\bibnamefont {Jiang}}, \bibinfo
  {author} {\bibfnamefont {B.}~\bibnamefont {Zhou}}, \bibinfo {author}
  {\bibfnamefont {Z.~J.}\ \bibnamefont {Wang}}, \bibinfo {author}
  {\bibfnamefont {Y.}~\bibnamefont {Zhang}}, \bibinfo {author} {\bibfnamefont
  {H.~M.}\ \bibnamefont {Weng}}, \bibinfo {author} {\bibfnamefont
  {D.}~\bibnamefont {Prabhakaran}}, \bibinfo {author} {\bibfnamefont {S.-K.}\
  \bibnamefont {Mo}}, \bibinfo {author} {\bibfnamefont {H.}~\bibnamefont
  {Peng}}, \bibinfo {author} {\bibfnamefont {P.}~\bibnamefont {Dudin}},
  \bibinfo {author} {\bibfnamefont {T.}~\bibnamefont {Kim}}, \bibinfo {author}
  {\bibfnamefont {M.}~\bibnamefont {Hoesch}}, \bibinfo {author} {\bibfnamefont
  {Z.}~\bibnamefont {Fang}}, \bibinfo {author} {\bibfnamefont {X.}~\bibnamefont
  {Dai}}, \bibinfo {author} {\bibfnamefont {Z.~X.}\ \bibnamefont {Shen}},
  \bibinfo {author} {\bibfnamefont {D.~L.}\ \bibnamefont {Feng}}, \bibinfo
  {author} {\bibfnamefont {Z.}~\bibnamefont {Hussain}},\ and\ \bibinfo {author}
  {\bibfnamefont {Y.~L.}\ \bibnamefont {Chen}},\ }\href
  {https://doi.org/10.1038/nmat3990} {\bibfield  {journal} {\bibinfo  {journal}
  {Nat. Mater.}\ }\textbf {\bibinfo {volume} {13}},\ \bibinfo {pages} {677}
  (\bibinfo {year} {2014})}\BibitemShut {NoStop}%
\bibitem [{\citenamefont {Borisenko}\ \emph {et~al.}(2014)\citenamefont
  {Borisenko}, \citenamefont {Gibson}, \citenamefont {Evtushinsky},
  \citenamefont {Zabolotnyy}, \citenamefont {B\"uchner},\ and\ \citenamefont
  {Cava}}]{PhysRevLett.113.027603}%
  \BibitemOpen
  \bibfield  {author} {\bibinfo {author} {\bibfnamefont {S.}~\bibnamefont
  {Borisenko}}, \bibinfo {author} {\bibfnamefont {Q.}~\bibnamefont {Gibson}},
  \bibinfo {author} {\bibfnamefont {D.}~\bibnamefont {Evtushinsky}}, \bibinfo
  {author} {\bibfnamefont {V.}~\bibnamefont {Zabolotnyy}}, \bibinfo {author}
  {\bibfnamefont {B.}~\bibnamefont {B\"uchner}},\ and\ \bibinfo {author}
  {\bibfnamefont {R.~J.}\ \bibnamefont {Cava}},\ }\href
  {https://doi.org/10.1103/PhysRevLett.113.027603} {\bibfield  {journal}
  {\bibinfo  {journal} {Phys. Rev. Lett.}\ }\textbf {\bibinfo {volume} {113}},\
  \bibinfo {pages} {027603} (\bibinfo {year} {2014})}\BibitemShut {NoStop}%
\bibitem [{\citenamefont {Jeon}\ \emph {et~al.}(2014)\citenamefont {Jeon},
  \citenamefont {Zhou}, \citenamefont {Gyenis}, \citenamefont {Feldman},
  \citenamefont {Kimchi}, \citenamefont {Potter}, \citenamefont {Gibson},
  \citenamefont {Cava}, \citenamefont {Vishwanath},\ and\ \citenamefont
  {Yazdani}}]{Jeon2014}%
  \BibitemOpen
  \bibfield  {author} {\bibinfo {author} {\bibfnamefont {S.}~\bibnamefont
  {Jeon}}, \bibinfo {author} {\bibfnamefont {B.~B.}\ \bibnamefont {Zhou}},
  \bibinfo {author} {\bibfnamefont {A.}~\bibnamefont {Gyenis}}, \bibinfo
  {author} {\bibfnamefont {B.~E.}\ \bibnamefont {Feldman}}, \bibinfo {author}
  {\bibfnamefont {I.}~\bibnamefont {Kimchi}}, \bibinfo {author} {\bibfnamefont
  {A.~C.}\ \bibnamefont {Potter}}, \bibinfo {author} {\bibfnamefont {Q.~D.}\
  \bibnamefont {Gibson}}, \bibinfo {author} {\bibfnamefont {R.~J.}\
  \bibnamefont {Cava}}, \bibinfo {author} {\bibfnamefont {A.}~\bibnamefont
  {Vishwanath}},\ and\ \bibinfo {author} {\bibfnamefont {A.}~\bibnamefont
  {Yazdani}},\ }\href {https://doi.org/10.1038/nmat4023} {\bibfield  {journal}
  {\bibinfo  {journal} {Nat. Mater.}\ }\textbf {\bibinfo {volume} {13}},\
  \bibinfo {pages} {851} (\bibinfo {year} {2014})}\BibitemShut {NoStop}%
\bibitem [{\citenamefont {Cano}\ \emph {et~al.}(2017)\citenamefont {Cano},
  \citenamefont {Bradlyn}, \citenamefont {Wang}, \citenamefont {Hirschberger},
  \citenamefont {Ong},\ and\ \citenamefont {Bernevig}}]{PhysRevB.95.161306}%
  \BibitemOpen
  \bibfield  {author} {\bibinfo {author} {\bibfnamefont {J.}~\bibnamefont
  {Cano}}, \bibinfo {author} {\bibfnamefont {B.}~\bibnamefont {Bradlyn}},
  \bibinfo {author} {\bibfnamefont {Z.}~\bibnamefont {Wang}}, \bibinfo {author}
  {\bibfnamefont {M.}~\bibnamefont {Hirschberger}}, \bibinfo {author}
  {\bibfnamefont {N.~P.}\ \bibnamefont {Ong}},\ and\ \bibinfo {author}
  {\bibfnamefont {B.~A.}\ \bibnamefont {Bernevig}},\ }\href
  {https://doi.org/10.1103/PhysRevB.95.161306} {\bibfield  {journal} {\bibinfo
  {journal} {Phys. Rev. B}\ }\textbf {\bibinfo {volume} {95}},\ \bibinfo
  {pages} {161306(R)} (\bibinfo {year} {2017})}\BibitemShut {NoStop}%
\bibitem [{\citenamefont {Murakami}(2006)}]{PhysRevLett.97.236805}%
  \BibitemOpen
  \bibfield  {author} {\bibinfo {author} {\bibfnamefont {S.}~\bibnamefont
  {Murakami}},\ }\href {https://doi.org/10.1103/PhysRevLett.97.236805}
  {\bibfield  {journal} {\bibinfo  {journal} {Phys. Rev. Lett.}\ }\textbf
  {\bibinfo {volume} {97}},\ \bibinfo {pages} {236805} (\bibinfo {year}
  {2006})}\BibitemShut {NoStop}%
\bibitem [{\citenamefont {Jensen}\ \emph {et~al.}(2013)\citenamefont {Jensen},
  \citenamefont {Kovtun},\ and\ \citenamefont {Ritz}}]{Jensen2013}%
  \BibitemOpen
  \bibfield  {author} {\bibinfo {author} {\bibfnamefont {K.}~\bibnamefont
  {Jensen}}, \bibinfo {author} {\bibfnamefont {P.}~\bibnamefont {Kovtun}},\
  and\ \bibinfo {author} {\bibfnamefont {A.}~\bibnamefont {Ritz}},\ }\href
  {https://doi.org/10.1007/JHEP10(2013)186} {\bibfield  {journal} {\bibinfo
  {journal} {J. High Energy Phys.}\ }\textbf {\bibinfo {volume} {2013}},\
  \bibinfo {pages} {186}}\BibitemShut {NoStop}%
\bibitem [{\citenamefont {Shi}\ \emph {et~al.}(2007)\citenamefont {Shi},
  \citenamefont {Vignale}, \citenamefont {Xiao},\ and\ \citenamefont
  {Niu}}]{PhysRevLett.99.197202}%
  \BibitemOpen
  \bibfield  {author} {\bibinfo {author} {\bibfnamefont {J.}~\bibnamefont
  {Shi}}, \bibinfo {author} {\bibfnamefont {G.}~\bibnamefont {Vignale}},
  \bibinfo {author} {\bibfnamefont {D.}~\bibnamefont {Xiao}},\ and\ \bibinfo
  {author} {\bibfnamefont {Q.}~\bibnamefont {Niu}},\ }\href
  {https://doi.org/10.1103/PhysRevLett.99.197202} {\bibfield  {journal}
  {\bibinfo  {journal} {Phys. Rev. Lett.}\ }\textbf {\bibinfo {volume} {99}},\
  \bibinfo {pages} {197202} (\bibinfo {year} {2007})}\BibitemShut {NoStop}%
\bibitem [{\citenamefont {Shitade}(2014)}]{Shitade01122014}%
  \BibitemOpen
  \bibfield  {author} {\bibinfo {author} {\bibfnamefont {A.}~\bibnamefont
  {Shitade}},\ }\href {https://doi.org/10.1093/ptep/ptu162} {\bibfield
  {journal} {\bibinfo  {journal} {Prog. Theor. Exp. Phys.}\ }\textbf {\bibinfo
  {volume} {2014}},\ \bibinfo {pages} {123I01} (\bibinfo {year}
  {2014})}\BibitemShut {NoStop}%
\bibitem [{\citenamefont {Nakai}\ and\ \citenamefont
  {Nomura}(2016)}]{PhysRevB.93.214434}%
  \BibitemOpen
  \bibfield  {author} {\bibinfo {author} {\bibfnamefont {R.}~\bibnamefont
  {Nakai}}\ and\ \bibinfo {author} {\bibfnamefont {K.}~\bibnamefont {Nomura}},\
  }\href {https://doi.org/10.1103/PhysRevB.93.214434} {\bibfield  {journal}
  {\bibinfo  {journal} {Phys. Rev. B}\ }\textbf {\bibinfo {volume} {93}},\
  \bibinfo {pages} {214434} (\bibinfo {year} {2016})}\BibitemShut {NoStop}%
\bibitem [{\citenamefont {Ominato}\ \emph {et~al.}(2019)\citenamefont
  {Ominato}, \citenamefont {Tatsumi},\ and\ \citenamefont
  {Nomura}}]{PhysRevB.99.085205}%
  \BibitemOpen
  \bibfield  {author} {\bibinfo {author} {\bibfnamefont {Y.}~\bibnamefont
  {Ominato}}, \bibinfo {author} {\bibfnamefont {S.}~\bibnamefont {Tatsumi}},\
  and\ \bibinfo {author} {\bibfnamefont {K.}~\bibnamefont {Nomura}},\ }\href
  {https://doi.org/10.1103/PhysRevB.99.085205} {\bibfield  {journal} {\bibinfo
  {journal} {Phys. Rev. B}\ }\textbf {\bibinfo {volume} {99}},\ \bibinfo
  {pages} {085205} (\bibinfo {year} {2019})}\BibitemShut {NoStop}%
\bibitem [{\citenamefont {Koshino}\ and\ \citenamefont
  {Hizbullah}(2016)}]{PhysRevB.93.045201}%
  \BibitemOpen
  \bibfield  {author} {\bibinfo {author} {\bibfnamefont {M.}~\bibnamefont
  {Koshino}}\ and\ \bibinfo {author} {\bibfnamefont {I.~F.}\ \bibnamefont
  {Hizbullah}},\ }\href {https://doi.org/10.1103/PhysRevB.93.045201} {\bibfield
   {journal} {\bibinfo  {journal} {Phys. Rev. B}\ }\textbf {\bibinfo {volume}
  {93}},\ \bibinfo {pages} {045201} (\bibinfo {year} {2016})}\BibitemShut
  {NoStop}%
\bibitem [{\citenamefont {Koshino}\ and\ \citenamefont
  {Hizbullah}(2019)}]{PhysRevB.99.209903}%
  \BibitemOpen
  \bibfield  {author} {\bibinfo {author} {\bibfnamefont {M.}~\bibnamefont
  {Koshino}}\ and\ \bibinfo {author} {\bibfnamefont {I.~F.}\ \bibnamefont
  {Hizbullah}},\ }\href {https://doi.org/10.1103/PhysRevB.99.209903} {\bibfield
   {journal} {\bibinfo  {journal} {Phys. Rev. B}\ }\textbf {\bibinfo {volume}
  {99}},\ \bibinfo {pages} {209903(E)} (\bibinfo {year} {2019})}\BibitemShut
  {NoStop}%
\bibitem [{\citenamefont {Bardeen}\ and\ \citenamefont
  {Zumino}(1984)}]{BARDEEN1984421}%
  \BibitemOpen
  \bibfield  {author} {\bibinfo {author} {\bibfnamefont {W.~A.}\ \bibnamefont
  {Bardeen}}\ and\ \bibinfo {author} {\bibfnamefont {B.}~\bibnamefont
  {Zumino}},\ }\href {https://doi.org/10.1016/0550-3213(84)90322-5} {\bibfield
  {journal} {\bibinfo  {journal} {Nucl. Phys. B}\ }\textbf {\bibinfo {volume}
  {244}},\ \bibinfo {pages} {421} (\bibinfo {year} {1984})}\BibitemShut
  {NoStop}%
\end{thebibliography}
\end{document}